\documentstyle[12pt,epsf]{article}
\begin{document}

\baselineskip 16pt plus 2pt minus 2pt

\newcommand{\sst}[1]{{\scriptscriptstyle #1}}
\newcommand{\beq}{\begin{equation}}
\newcommand{\eeq}{\end{equation}}
\newcommand{\beqa}{\begin{eqnarray}}
\newcommand{\eeqa}{\end{eqnarray}}
\newcommand{\dida}[1]{/ \!\!\! #1}
\renewcommand{\Im}{\mbox{\sl{Im}}}
\renewcommand{\Re}{\mbox{\sl{Re}}}
\def\simge{\hspace*{0.2em}\raisebox{0.5ex}{$>$}
     \hspace{-0.8em}\raisebox{-0.3em}{$\sim$}\hspace*{0.2em}}
\def\simle{\hspace*{0.2em}\raisebox{0.5ex}{$<$}
     \hspace{-0.8em}\raisebox{-0.3em}{$\sim$}\hspace*{0.2em}}
\def\GAS{{G_{\sst{A}}^s}}
\def\GES{{G_{\sst{E}}^s}}
\def\GMS{{G_{\sst{M}}^s}}
\def\GiS{{G_{\sst{i}}^s}}
\def\GAis{{G_{\sst{A}}^\sst{(I=0)}}}
\def\GEis{{G_{\sst{E}}^\sst{(I=0)}}}
\def\GMis{{G_{\sst{M}}^\sst{(I=0)}}}
\def\FIis{{F_i^\sst{(I=0)}}}
\def\FOis{{F_1^\sst{(I=0)}}}
\def\FTis{{F_2^\sst{(I=0)}}}
\def\FIS{{F_i^s}}
\def\FOS{{F_1^s}}
\def\FTS{{F_2^s}}
\def\GA{{G_{\sst{A}}}}
\def\GE{{G_{\sst{E}}}}
\def\GM{{G_{\sst{M}}}}
\def\FI{{F_i}}
\def\FO{{F_1}}
\def\FT{{F_2}}
\def\bra#1{{\langle#1\vert}}
\def\ket#1{{\vert#1\rangle}}
\def\pbar{{\bar{p}}}
\def\notder{{\not\! \partial}}
\def\mn{{m_{\sst{N}}}}
\def\mns{{m^2_{\sst{N}}}}
\def\mks{{m_{\sst{K}}^2}}
\def\mk{{m_{\sst{K}}}}
\def\fkem{{F_{\sst{K}}^{\sst{EM}}}}
\def\fks{{F_{\sst{K}}^s}}
\def\fka{{F_{\sst{K}}^a}}
\def\mpi{{m_\pi}}
\def\mpis{{m_\pi^2}}
\def\bpp{{b_1^{1/2,\,1/2}}}
\def\bpm{{b_1^{1/2,\,-1/2}}}
\def\bppm{{b_1^{1/2,\,\pm 1/2}}}
\def\bll{{b_1^{\lambda, \bar{\lambda}}}}
\def\sgmus{{\bar{s}\gamma_\mu s}}
\def\Kbar{{\bar K}}
\def\sbar{{\bar s}}
\def\FOa{{F_1^{(a)}}}
\def\FTa{{F_2^{(a)}}}
\def\GEa{{G_\sst{E}^{(a)}}}
\def\GMa{{G_\sst{M}^{(a)}}}

\begin{titlepage}


\hfill{INT \#DOE/ER/40561-18-INT98}

\hfill{TRI-PP-98-26}

\vspace{1.0cm}

\begin{center}
{\large {\bf Spectral Content of Isoscalar Nucleon Form Factors}}

\vspace{1.2cm}

H.-W. Hammer$^{a,}$\footnote{email: hammer@triumf.ca} and
M.J. Ramsey-Musolf$^{b,c,}$\footnote{email: mjrm@phys.uconn.edu}

\vspace{0.8cm}

$^a$ TRIUMF, 4004 Wesbrook Mall, Vancouver, BC, Canada V6T 2A3\\
$^b$ Department of Physics, University of Connecticut, Storrs, CT
06269, USA\\
$^c$ Institute for Nuclear Theory,
University of Washington, Seattle, WA 98195, USA\\[0.4cm]
\end{center}

\vspace{1cm}

\begin{abstract}
The nucleon strange vector and isoscalar
electromagnetic form factors are studied using a spectral decomposition.
The $K\bar{K}$ contribution to the electric and magnetic radii as well as
the magnetic moment is evaluated to all orders in the strong interaction
using an analytic continuation of experimental $KN$ scattering amplitudes
and bounds from unitarity. The relationship
between non-resonant and resonant $K\bar{K}$ contributions to the
form factors is demonstrated, and values for the vector and tensor
$\phi N\bar{N}$ couplings are derived. The $K\bar{K}$ spectral functions are
used to evaluate the credibility of model calculations for the strange
quark vector current form factors.\\[0.3cm]
\end{abstract}

\vspace{2cm}
\vfill
\end{titlepage}

\section{Introduction}
\label{sec:intro}
The reasons for the success of the constituent quark model of light hadrons
remains one of the on-going mysteries of strong interaction physics.
Although deep inelastic scattering has provided incontrovertible
evidence for the existence of gluons and QCD current quarks in the
lightest hadrons, these degrees or freedom are manifestly absent from the
the quark model. Nevertheless, a description of
light hadrons solely in terms of
constituent quarks moving in an effective potential has been enormously
successful in accounting for the mass spectrum and other properties of
low-lying hadrons. Various explanations for this situation have appeared
in the literature, including the simple and intuitive idea that the sea
quarks and gluons of QCD \lq\lq renormalize" the valence current quarks
into the constituent quarks of the quark model \cite{KaM88}.
In this picture, for
example, the multitude of QCD degrees of freedom appear to a long
wavelength probe primarily as single objects carrying the quantum
numbers and effective mass of the constituent quark. From the standpoint
of the quark-quark effective potential, gluon and sea-quarks are
similarly un-discernible -- as they help renormalize the quark model string
tension into the physical value used as model input \cite{GeI90}.
In fact, most low-energy observables studied to date are unable to
uncover explicit signatures of QCD degrees of freedom.

There have been, however, a few exceptions to this situation. Of
particular interest are observables sensitive to the presence
of strange quarks in the nucleon. In contrast to up- and down-quarks,
which appear both as valence and sea quarks, strange quarks constitute
a purely sea-quark degree of freedom. Being the lightest such objects,
they ought to generate the largest effects (in comparison to the heavier
quarks). Consequently, nucleon matrix elements of strange quark operators
provide an interesting window on the $q\bar q$ sea and as such may shed
new light on the connection between non-perturbative QCD and the quark model.
Indeed, were strange-quark observables found to be vanishingly small,
one might ascribe the quark model's success partly to the numerical
insignificance of sea quark effects.\footnote{The reason {\it why}
non-perturbative QCD produces small sea-quark effects at low-energies
would remain to be explained, however.} In fact, the situation is more
ambiguous.
As is well-known, analyses of the \lq\lq $\sigma$-term" in $\pi N$
scattering, the $g_1$ sum in polarized deep inelastic scattering, and
$\nu_\mu\ (\bar{\nu}_\mu)\ N$ deep inelastic scattering suggest that
non-trivial fractions of the nucleon mass, spin, and light-cone momentum
arise from the $s\bar{s}$ sea (see Ref. \cite{Mus94} and references 
therein). Evidently, the most na\"\i ve
explanation for the quark model's validity is ruled out by these
analyses.

More recently, a well-defined program has begun to determine the matrix
element $\bra{N}\sbar\gamma_\mu s\ket{N}$ using parity-violating elastic
and quasi-elastic electron scattering from the proton and
nuclei \cite{Mus94}. The first result for the
magnetic form factor associated with this matrix element has been
reported by the SAMPLE collaboration at
MIT-Bates \cite{Mue97}:
\beq
\label{intro:sample}
\GMS(q^2 =-0.1\mbox{ GeV}^2)=0.23\pm 0.37\pm 0.15\pm 0.19\,,
\eeq
where $q^2$ is the four-momentum transfer squared.
(The first error is statistical, the second is the estimated systematic
error, and the last uncertainty is due to radiative corrections entering
the analysis.) Although the value is consistent with zero, the error
bars are large. Improved accuracy is expected when the full data set
is analyzed. Similarly, a combination of the strange magnetic and
electric form factors have been determined by the HAPPEX
collaboration\cite{happex}:
\beq
\label{intro:happex}
\GES+0.39\GMS(q^2 =-0.48\mbox{ GeV}^2)=0.023\pm 0.034\pm 0.022\pm 0.026\,,
\eeq
where the first two errors are again of statistical and systematic origin,
respectively, and the last one arises from the estimated uncertainty in the
electric neutron form factor.
While no definitive conclusion can as yet be made regarding the
experimental scale of $\bra{N}\sbar\gamma_\mu s\ket{N}$, one expects
to be able to do so at the conclusion of the measurements.

In contrast, the theoretical understanding of $\bra{N}\sbar\gamma_\mu
s\ket{N}$ is much less clear. The difficulty lies with the mass scales
relevant to strange quark dynamics. In contrast
to the heavy quarks, for which $m_q \gg \Lambda_{QCD}$, the strange
quark has $m_s\sim \Lambda_{QCD}$. Consequently, the lifetime of
a virtual $s\sbar$ pair is commensurate with typical strong interaction
timescales, allowing the pair to exchange a plethora of gluons with
other quarks and gluons in its environment. The dynamics of the pair are
therefore inherently non-perturbative. Given the present state of
QCD theory, a complete, first principles treatment of
$\bra{N}\sbar\gamma_\mu s\ket{N}$ has remained beyond reach. Attempts to
obtain this matrix element on the lattice have produced two results
for $\GMS(q^2=0)$, neither of which agree with each other nor with
the first SAMPLE results \cite{Lei96,DLW98}, and one result for the
slope of $\GES$ at the origin with large error bars \cite{DLW98}.

An alternative -- and more popular approach -- has been to employ
various effective frameworks, with varying degrees
of model-dependence. These frameworks have included nucleon
models, chiral perturbation theory (ChPT), and dispersion
relations. Generally speaking, the degrees of freedom adopted in each
of these approaches have been hadronic rather than quark and gluon, given
that the lifetime of an $s\sbar$ pair permits it to form strange hadronic
states. Apart from a few exceptions, effective approaches do not address
the way in which QCD sea quarks hadronize. Hence, the connection with
QCD is indirect at best, with each approach emphasizing some aspects of
the strong interaction to the exclusion of others.

Not surprisingly, the range of predictions for the strangeness
form factors is broad. In particular, the breadth of {\em model}
predictions appears to be as wide as the variety of models that has been
used even though the same models are in reasonable agreement for standard
nucleon observables \cite{Mus94,MuB94,GeI97}. This situation illustrates the
sensitivity of sea quark observables to model assumptions and the limited
usefulness of models in making airtight predictions.
One might have hoped for more insight from ChPT,
which relies on the chiral symmetry of QCD to successfully account
for a wide variety of other low-energy observables \cite{BKM95}.
Unfortunately,
ChPT is unable to make a prediction for the leading non-vanishing parts
of $\GMS$ or $\GES$ since the leading moments depend on unknown
counterterms
\cite{MuI97}. Recently, however, it has been noticed that slope of
$\GMS$ at the origin is independent of unknown counterterms to
${\cal O}(p^3)$ \cite{HMS98}.

In the present study, we turn to dispersion relations (DR's) to
derive insight into $\bra{N}\sbar\gamma_\mu s\ket{N}$. Like
ChPT, DR's rely on certain general features of QCD
(and other field theories) to relate existing experimental
data to the observables of interest. In the case of DR's it
is analyticity and causality, rather than chiral symmetry, which allow
one to make the connection. Although DR's do not bear on the way in
which QCD quarks and gluons form intermediate strange hadronic states,
they do provide an essentially model-independent framework for treating
the way in which those states contribute to the form factors. We view
them as providing an intermediate step toward understanding the
strange quark form factors at the fundamental level of QCD. Because
of their generality, they also allow us to evaluate the credibility
of several model predictions.

Our use of DR's to study nucleon form factors is not new. The
spectral content of the nucleon isovector form factors has been 
clearly delineated using this approach \cite{Hoe75,MMD96}. It is now
known that both an un-correlated $\pi\pi$ continuum as well as the
$\pi\pi\to\rho$ resonance play important roles in the low-$q^2$ behavior
of these form factors. The $q^2$-dependence of the isoscalar EM form
factors has been successfully reproduced using DR's under the assumption
of vector meson dominance (VMD). The results have been used to infer
relations between the
$\omega NN$ and $\phi NN$ coupling strengths and to make predictions for
the strange quark form factors \cite{Jaf89,HMD96,For96}. However, the
relationship between the resonance and continuum contributions to these
form factors has not been previously established. Consequently, a number
of model predictions have appeared which rely on the assumption that the
uncorrelated continuum (\lq\lq meson cloud") gives the largest effect.
These meson cloud calculations have generally entailed a truncation at
second order in the strong hadronic coupling, $g$ -- a practice of
questionable validity. The corresponding predictions have generally
been in disagreement with those obtained using  VMD.

In what follows, we consider both the strange quark and isoscalar
EM form factors without relying on the a priori assumption of
vector meson or meson cloud dominance.  We focus in particular on the
contribution from the $K\Kbar$ intermediate state. The rationale for
this focus is twofold. First, the $K\Kbar$ state constitutes the lightest
intermediate state containing valence $s$ and $\sbar$ quarks. Its
contribution to the strange quark form factors has correspondingly
been emphasized in both models and ChPT.
Second, given the present availability of strong interaction and
EM data, the $K\bar{K}$ contribution can be computed to all orders in
$g$ using a minimum of assumptions.
From an analysis of $KN\to K N$ and $e^+e^-\to K\Kbar$
data, we show that the scale of the $K\Kbar$ contribution depends
critically on effects going beyond ${\cal O}(g^2)$ and argue that a
similar situation holds for the remainder of the form factor spectral
content. We also

\medskip
\noindent (a) illustrate the relation between the continuum and resonance
contributions,

\medskip
\noindent (b) evaluate the credibility of several model predictions as
well as the ${\cal O}(p^3)$ prediction of ChPT for the magnetic radius,

\medskip
\noindent (c) derive values for the vector and tensor $\phi NN$
couplings and compare with those obtained from isoscalar EM form
factors under the assumption of VMD.

\medskip
In Refs. \cite{MHD97,RMH98}, we reported on the results of our DR
analysis of the $K\bar{K}$ contribution to the nucleon
\lq\lq strangeness radius" (the slope of $\GES$ at the photon point).
Here, we expand on that analysis to consider the full $q^2$-dependence in
both the isoscalar EM and strangeness channels and to discuss both the
electric and magnetic form factors. Since the DR approach requires knowledge
of the $K\Kbar \to N{\bar N}$ amplitudes in the unphysical region, some
form of analytic continuation is needed to complete the analysis.
Using backward dispersion relations, we obtain the unphysical amplitudes
from $K N$ phase shift analyses. The results of this continuation and their
implications for nucleon form factors constitute a central theme of this
paper.

Our discussion of these issues is
organized as follows. After outlining our formalism, we perform the
spectral decomposition of the form factors and write down DR's in Section
\ref{sec:drsd}. In Section \ref{sec:kkbar}, we express the spectral
functions in terms of $K\bar{K} \to N\bar{N}$ partial waves and
give the corresponding unitarity bounds valid in the physical
region of the dispersion integrals. The analytic continuation of $KN$
amplitudes which is used in the unphysical region is performed in
Section \ref{sec:anacon}. A brief description of the analytic 
continuation and our treatment of the inherent problems is given.
The results are applied to the nucleon's strange and isoscalar EM
form factors in Section \ref{sec:kkbarnucff}. In Section
\ref{sec:conc}, we discuss the contribution of other intermediate
states and summarize our conclusions.

\section{Spectral Decomposition and Dispersion Relations}
\label{sec:drsd}
The vector current form factors of the nucleon, $F_1(t)$ and $F_2(t)$, 
are defined by
\beq
\bra{N(p')} j_\mu(0) \ket{N(p)}=\bar{u}(p')\left[
F_1(t)\gamma_\mu+{iF_2(t)\over 2\mn}\sigma_{\mu\nu}(p'-p)^\nu
\right]u(p)\, .
\eeq
where $t=q^2=(p'-p)^2$. We consider two cases for $j_\mu$: (i) the
strange vector current $\bar{s} \gamma_\mu s$ and (ii) the isoscalar
EM current $j_\mu^\sst{(I=0)}$.
Since the nucleon carries no net strangeness, $\FOS$ must vanish at zero
momentum transfer, (i.e. $\FOS(0)=0$), whereas $\FOis$ is normalized
to the isoscalar EM charge of the nucleon, $\FOis(0)=1/2$.
We also define the electric and magnetic Sachs form factors,
which may be interpreted as the fourier transforms of the charge and
magnetic moment distributions in the Breit frame,
\begin{eqnarray}
\label{drsd:dsacsff}
& &\GE=\FO-\tau\FT\,,\qquad\qquad\qquad
\GM=\FO+\FT\,,
\end{eqnarray}
with $\tau=-t/4\mns$. In the case of the strange form factors 
we are particularly interested in their leading moments, the strange 
magnetic moment and the strange radii:
\beqa
\label{drsd:moms}
& &\kappa^s = \FTS(0) = \GMS(0)=\mu^s\,,
\vphantom{\frac{1}{2}}\\
& &\langle r^2\rangle^s_i = 6{d\GiS(t)\over d t}\bigg\vert_{t=0}\,,
\nonumber
\eeqa
where $i=E,M$, respectively. A dimensionless version of the
radii can be defined by
\begin{eqnarray}
\rho^s_E&=&{d\GES(\tau)\over d\tau}\bigg\vert_{\tau=0}=
-\frac{2}{3}\mns\langle r^2\rangle^s_E 
 \label{drsd:rhos}
\end{eqnarray}
and similarly for $\rho^s_M$.
The leading moments of the EM form factors are defined analogously. 

It is conventional to employ a once subtracted DR for
$F_1$. Typically, one wishes to predict $F_2(0)$ as well as its
$t\not= 0$ behavior. In this case, an unsubtracted DR
is appropriate. We follow this ansatz in the present
study and obtain
\beq
\label{drsd:fts}
\FTS(t)=\frac{1}{\pi}\int_{t_\lambda}^{\infty}
\frac{\Im\,\FTS(t')}{t'-t}dt' \,,
\eeq
and
\beq
\label{drsd:fos}
\FOS(t)=\frac{t}{\pi}\int_{t_\lambda}^{\infty}
\frac{\Im\,\FOS(t')}{t'(t'-t)}dt' \, .
\eeq
As a consequence, subtracted DR's can be written
for the Sachs form factors as well. 

From Eqs. (\ref{drsd:fts}-\ref{drsd:fos}), it is clear that the
quantities of interest are the imaginary parts of the form factors. The
success of the DR analysis relies on a decomposition of the ${\Im\, F_i}$
into scattering amplitudes involving physical states. To obtain this
spectral decomposition, we follow the treatment of Refs.
\cite{FGT58,DrZ60,MHD97} and consider the crossed matrix element
\beq
\label{drsd:defjm}
J_\mu=\bra{N(p)\,\bar{N}(\bar{p})} j_\mu(0) \ket{0}=\bar{u}(p)
\left[ F_1(t)\gamma_\mu+{iF_2(t)\over 2\mn}\sigma_{\mu\nu}
(\bar{p}+p)^\nu \right]v(\bar{p})\, ,
\eeq
where $t$ is now timelike.
Using the LSZ reduction formalism and inserting a complete set
of intermediate states, $\Im\,J_\mu$ may be expressed as
\beqa
\Im\,J_\mu &=&\frac{\pi}{\sqrt{Z}}(2\pi)^{3/2}{\cal N}
\sum_\lambda \bra{N(p)}\bar{J_N}(0)\ket{\lambda}\bra{\lambda}
j_\mu(0) \ket{0}\,v(\bar{p})\,\delta^4(p+\bar{p}-p_\lambda)\, ,
\label{drsd:ImJ}
\eeqa
where ${\cal N}$ is a spinor normalization factor and $J_N(0)$
a nucleon source. Eq. (\ref{drsd:ImJ}) determines the singularity 
structure of the form factors and relates their imaginary parts 
to on-shell matrix elements for other processes.
The form factors have multiple cuts on the
positive real $t$-axis. The invariant
mass-squared $M_\lambda^2$ of the lightest state appearing in the 
sum defines the beginning of the first cut and the lower limit in the 
dispersion integrals: $M_\lambda^2=t_\lambda$.
Since Eq. (\ref{drsd:ImJ}) is linear, the contributions of
different $\ket{\lambda}$ can be treated separately.

There is an infinite number of contributing intermediate states
$\ket{\lambda}$ which are restricted by the quantum numbers of
the currents $\bar{s}\gamma_\mu s$ and $j_\mu^\sst{(I=0)}$
$[I^G (J^{PC}) = 0^- (1^{--})]$. Na\"\i vely, the lightest states
generate the most important contributions to the leading moments of
the current. Moreover, because of the 
source $J_N(0)$, the intermediate states must have zero baryon number.
The lowest allowed states together with their thresholds
are collected in Table \ref{drsd:untab1a}.
Resonances, such as the
$\omega$, do not correspond to asymptotic states and are already included
in the continuum contributions, such as that from the $3\pi$ state.
\begin{table}[htb]
\begin{center}
\begin{tabular}{|c|c||c|c|}
\hline mesonic states & $t_\lambda [\,\mbox{GeV}^2]$ & baryonic states &
$t_\lambda [\,\mbox{GeV}^2]\vphantom{\Large[}$ \\
\hline\hline 
$3\pi$ & 0.18 & $N\bar{N}$ & 3.53 \\
$5\pi$ & 0.49 & $N\bar{N}\pi\pi$ & 4.67 \\
$7\pi$ & 0.96 & $\Lambda\bar{\Lambda}$ & 4.84 \\
$K\bar{K}$ & 0.98 & $\Sigma\bar{\Sigma}$ & 5.76 \\
$K\bar{K}\pi$ & 1.28 & $\Lambda\bar{\Sigma}\pi$ & 5.95 \\
$\vdots$ && $\vdots$ & \\ \hline
\end{tabular}
\end{center}
\caption{\label{drsd:untab1a}Lowest mass intermediate states 
contributing to Eq. (\protect\ref{drsd:ImJ}).}
\end{table}

When sufficient 
data exist, experimental information may be used to determine
the matrix elements appearing in Eq. (\ref{drsd:ImJ}). 
However, when the threshold $t_\lambda$ of the intermediate state
$\ket{\lambda}$ is below the two-nucleon threshold, the values of
the matrix element $\bra{N(p)}\bar{J_N}(0)\ket{\lambda}\,v(\bar{p})$
are also required in the unphysical region $t_\lambda
\leq t \leq 4\,\mns$. In this case, the amplitude must be analytically
continued from the physical to the unphysical regime.
The first cut in the complex $t$-plane
appears at the $3\pi$ production threshold, $t=9m_\pi^2$, and
higher-mass intermediate states generate additional cuts.
For example for $\ket{\lambda}=\ket{K \bar{K}}$
the cut runs from $t=4\mks$ to infinity. Therefore, the matrix element
for $K\bar{K}\longrightarrow\bar{N}N$ is also needed in the unphysical
region $4\mks\leq t\leq 4\mns$, which requires an
analytic continuation. 

Some of the predictions for the $\FIS$ reported in the literature are
based on approximations to the spectral functions appearing in Eqs.
(\ref{drsd:fts}, \ref{drsd:fos}). The work of Refs. 
\cite{Jaf89,HMD96,For96} employed a VMD approximation, which amounts
to writing the spectral function as 
$\Im\,  \FIS(t) = \pi\sum_j a_j\delta(t-m_j^2)$,
where \lq\lq $j$" denotes a particular vector meson resonance
(e.g. $\omega$ or $\phi$) and the sum runs over a finite number of
resonances. In terms of the formalism from above
this approximation omits any explicit
multi-meson intermediate states $\ket{n}$ and assumes that
the products $\bra{N(p)}\bar{J_N}(0)\ket{\lambda}\bra{\lambda}
\bar{s}\gamma_\mu s\ket{0}v(\pbar)$ are strongly peaked
near the vector meson masses. The same has been made conventional 
analyses of the isoscalar EM form factors \cite{Hoe75,MMD96}.

In contrast, a variety of hadronic effective theory and
model calculations for the strange form factors have focused
on contributions from the two-kaon intermediate state
\cite{Mus94,MuB94,MuI97}. Even though $\ket{K\Kbar}$ is not the lightest
state appearing in Table \ref{drsd:untab1a}, it is the lightest
state containing valence strange quarks. The rationale for focusing
on the $\bar{K\Kbar}$ contribution is based primarily
on the intuition that such states
ought to give larger contributions to the matrix element $\bra{\lambda}
\bar{s}\gamma_\mu s\ket{0}$ than purely pionic states with no
valence $s$ or $\bar{s}$ quarks. In other words, the kaons represent
the lightest contribution favored by the OZI rule.
Typically, kaon-cloud predictions have been computed to ${\cal O}(g^2)$
only. The results for for $\rho^s_E$ in particular are smaller in magnitude
than the vector meson dominance predictions and have
the opposite sign. In what follows, we illustrate how both the structure
and magnitude of the full kaon cloud contribution differ
from the ${\cal O}(g^2)$ result and how a $\phi$-resonance structure appears
in the all-orders analysis.

Although we consider here only the $K\Kbar$ intermediate state, we note
in passing that the validity of this so called \lq\lq kaon cloud
dominance'' ansatz is open to question for a variety reasons.
As can be seen from Table \ref{drsd:untab1a}, for example,
the three-pion threshold is significantly below the $K\bar{K}$ threshold.
Consequently, the $3\pi$ contribution is weighted more strongly
in the dispersion integral than the $K\bar{K}$ contribution
because of the denominators in Eqs. (\ref{drsd:fts}, \ref{drsd:fos}).
Moreover, three pions can resonate into a state having the same
quantum numbers as the $\phi$ (nearly pure $s\bar{s}$), and thereby
generate a non-trivial contribution to the current matrix element.
Indeed, the $\phi$ has roughly a 15\% branch to multi-pion
final states (largely via a $\rho\pi$ resonance). Although such
resonances do not appear explicitly in the sum over the states in
Eq. (\ref{drsd:ImJ}), their impact nevertheless
enters via the current matrix element $\bra{3\pi}\bar{s}
\gamma_\mu s\ket{0}$ and the $N\bar{N}$ production amplitude
$\bra{N(p)}\bar{J_N}(0)\ket{3\pi}v(\pbar)$. Thus, the $3\pi$ state
could contribute appreciably to the strangeness form factors via its
coupling to the $\phi$. We return to this possibility in Section
\ref{sec:conc} (see also Ref. \cite{HaM98}).

\section{$K\bar{K}$ Intermediate State and Unitarity}
\label{sec:kkbar}
In order to determine $K\bar{K}$ contribution to the spectral
functions, we need the matrix elements
$\bra{N(p)}\bar{J_N}\ket{K(k)\bar{K}(\bar{k})}\,v(\bar{p})$ and
$\bra{K(k)\bar{K}(\bar{k})}j_\mu\ket{0}$.
By expanding the $K\bar{K}\to N\bar{N}$ amplitude
in partial waves, we are able to impose the constraints of unitarity 
in a straightforward way. In doing so, we follow the helicity
amplitude formalism of Jacob and Wick \cite{JaW59}.
With $\lambda$ and $\bar{\lambda}$ being the nucleon and antinucleon
helicities, we write the corresponding $S$-matrix element as
\begin{eqnarray}
\label{kkbar:s_norm}
& &\bra{N(p,\lambda)\bar{N}(\pbar,\bar{\lambda})}{\hat S}\ket{K(k)\bar{K}
(\bar{k})}= \\ & &\hphantom{N(p,\lambda)}
 i(2\pi)^4\delta^4(p+\pbar-k-\bar{k})(2\pi)^2 \left[{64 t\over t-4 m_K^2}
\right]^{1/2} \bra{\theta,\phi,\lambda,\bar{\lambda}}{\hat S}(P)\ket{00}
\,,\nonumber
\end{eqnarray}
where $t=P^2=(p+\pbar)^2$ and $m_K$ is the kaon mass. The matrix element
$\bra{\theta, \phi, \lambda, \bar{\lambda}}{\hat S}(P)\ket{00}$
is then expanded in partial waves as \cite{MHD97,JaW59}
\begin{equation}
\label{kkbar:par_d}
S_{\lambda, \bar{\lambda}} \equiv
\bra{\theta, \phi, \lambda, \bar{\lambda}}{\hat S}(P)\ket{00}=\sum_J
\left({2J+1\over 4\pi}\right) b_J^{\lambda, \bar{\lambda}} \;
{\cal D}_{0\mu}^J(\phi, \theta, -\phi)^{\ast}\,,
\end{equation}
where ${\cal D}_{\nu\, \nu'}^J(\alpha, \beta, \gamma)$ is a
Wigner rotation matrix with $\mu=\lambda-\bar{\lambda}$.
The $b_J^{\lambda\, \bar{\lambda}}$ define partial waves of
angular momentum $J$. Because of the quantum numbers of
the isoscalar EM and strange vector currents, only
the $J=1$ partial waves contribute to the
spectral functions. Moreover, because of 
parity invariance only two of the four partial 
waves are independent. We choose $\bpp$ and $\bpm$ which fulfill the 
threshold relation \cite{MHD97}
\begin{equation}
\label{kkbar:th_rel}
\left.\bpm(t)\right|_{t=4\mns}=\sqrt{2}\,\left.\bpp(t)\right|_{t=4\mns}\,.
\end{equation}
Using the above definitions, the unitarity of the $S$-matrix,
$S^{\dag} S = 1$, requires that
\begin{equation}
\label{kkbar:ubs}
|b_J^{\lambda, \bar{\lambda}}(t)|\leq 1 \,,
\end{equation}
for $t\geq 4\mns$.
Consequently, unitarity gives model-independent bounds on the
contribution of the physical region ($t\geq4\mns$) to the
imaginary part. In the unphysical region ($4\mks \leq t \leq 4\mns$),
however, the partial waves are not bounded by unitarity.
Therefore, we must rely upon an analytic continuation of
$KN$ scattering amplitudes. This procedure is
discussed in the next section. 

The second matrix element appearing in Eq. (\ref{drsd:ImJ}), 
$\bra{K(k)\bar{K}(\bar{k})} j^{(a)}_\mu \ket{0}$, is parametrized 
by the kaon vector current form factor $\fka$:
\begin{eqnarray}
\bra{0}j_\mu^{(a)}\ket{K(k) \bar{K}(\bar{k})}&=&(k-\bar{k})_\mu
\fka(t)\,,
\end{eqnarray}
where $a$ denotes $EM$ or $s$ and $\fka(0)$ gives the corresponding
charge (e.g., $\fks(0)=-1$). 

From Eq. (\ref{drsd:ImJ}), the spectral functions are related
to the partial waves and the kaon strangeness form factor by,
\begin{eqnarray}
\label{drsd:imf1}
\Im\,  \FOa(t)&=&\Re\,\left\{\left({\mn q_t\over 4 p_t^2}\right)\left[
{E\over\sqrt{2}\mn}\bpm(t)-\bpp(t)\right]\fka(t)^{\ast}\right\}\,,\\
&& \nonumber \\
\label{drsd:imf2}
\Im\,  \FTa(t)&=&\Re\,\left\{\left({\mn q_t\over 4 p_t^2}\right)\left[
\bpp(t)-{\mn\over\sqrt{2}E}\bpm(t)\right]\fka(t)^{\ast}\right\}\,,
\end{eqnarray}
where
\beq
\label{drsd:defqp}
p_t=\sqrt{t/4-\mns}\,,\qquad q_t=\sqrt{t/4-m_K^2}\,,\qquad
\mbox{and}\qquad E=\sqrt{t}/2\,.
\eeq
The corresponding spectral functions 
for the Sachs form factors follow from Eq. (\ref{drsd:dsacsff}),
\begin{eqnarray}
\label{drsd:imge}
\Im\,  \GEa(t)&=&\Re\,\left\{\left({q_t\over 4 \mn}\right)\bpp(t)
\fka(t)^{\ast}\right\}\,,\\
\label{drsd:imgm}
\Im\,  \GMa(t)&=&\Re\,\left\{\left({q_t\over 2 \sqrt{2t}}\right)
\bpm(t)\fka(t)^{\ast}\right\}\,.
\end{eqnarray}

On one hand,
Eqs. (\ref{drsd:imf1}-\ref{drsd:imgm}) may be used to determine
the spectral functions from experimental data.
On the other hand, one can impose bounds on the imaginary parts
in the physical region by using Eq. (\ref{kkbar:ubs}).
Eqs. (\ref{drsd:imf1}-\ref{drsd:imgm}) involve expressions
of the type
\begin{equation}
\label{uni:rel_p}
\Re\,\left\{\bppm(\fka)^{\ast}\right\}
=|\bppm| |\fka| \cos(\delta_1-\delta_K)=|\bppm| |\fka| (1+\gamma_K)\,,
\end{equation}
where the phase correction $\gamma_K$ is defined
by $\gamma_K\equiv\cos(\delta_1-\delta_K)-1$, with $\delta_1$
and $\delta_K$ the complex phases of the $\bll$
and form factor, respectively. 
The experimental information on $\gamma_K$ is incomplete.
Since $|1+\gamma_K|\leq 1$, however, we can take $\gamma_K=0$
to obtain an upper bound on the spectral functions.
In order to obtain finite bounds for the Dirac and Pauli form factors
at the $N\bar{N}$ threshold, we build in the
correct threshold relation for the $\bll$, Eq. (\ref{kkbar:th_rel}). 
This is necessary to cancel the $1/p_t^2$
factor in Eqs. (\ref{drsd:imf1}, \ref{drsd:imf2}). Strictly speaking,
the relation holds only for $t=4\mns$. For simplicity, however,
we assume this relation to be valid for all momentum transfers,
as e.g. holds in the tree approximation of perturbation theory.
Consequently, we have
\begin{eqnarray}
\label{kkbar:ubf1th}
|\Im\, \FOa(t)| &\leq& \frac{q_t}{2\sqrt{2} ( \sqrt{t} +2 \mn)}|\fka(t)|
\, ,\\
\label{kkbar:ubf2th}
|\Im\, \FTa(t)| &\leq& \frac{ \mn q_t}{\sqrt{2 t} (\sqrt{t} +2 \mn)}
|\fka(t)| \, .
\end{eqnarray}
The unitarity bounds for the Sachs form
factors are obtained more straightforwardly by simply setting
$|\bppm(t)| \leq 1$ in Eqs. (\ref{drsd:imge}, \ref{drsd:imgm}).

\subsection{Kaon Form Factors}
\label{sec:kffs}
The timelike EM kaon form factor, $\fkem(t)$, has been determined
from $e^+e^-\to K\bar{K}$ cross sections. A
striking feature of $\fkem$ observed in these studies is the pronounced
peak for $t\approx m_\phi^2$ \cite{Del81}. At higher values of $t$, 
oscillations at a much smaller scale are observed. A variety of analyses
of $\fkem$ in this region have been performed \cite{Del81,FeS81}, 
and it is found that $\fkem$ is well-described 
by a VMD parametrization:
\begin{equation}
\label{kff:vmd}
\fkem(t)=\sum_V C_V {m_V^2\over m_V^2-t-im_V\Gamma_V f_V(t)}\,,
\end{equation}
where the sum is over vector mesons of mass $m_V$ and width $\Gamma_V$
and where $f_V(t)$ is some specified function of $t$. 
We use $f_V(t)= t/m_V^2$ \cite{FeS81}.
In nearly all analyses, one finds for the residues: $C_\rho\approx 1/2$, 
$C_\omega\approx 1/6$, and $C_\phi\approx 1/3$. These residues may 
alternately be described in terms of the $V\gamma$ and $VK\bar{K}$ 
couplings: $C_V = g_{VK\bar{K}}/f_V\,,$ where 
$f_\rho\approx 5.1$, $f_\omega\approx 17$, and $f_\phi\approx 13$. The
strong couplings can be determined from $\Gamma(\phi\to K\bar{K})$ and
SU(3) relations. 

To obtain $\fks(t)$ we follow Refs. \cite{Jaf89,MuI97,HaM98} 
and draw upon the known flavor content of the vector mesons. 
The $\rho$ does not contribute to isoscalar form factors.
To the extent that (i) the $\omega$ and
$\phi$ satisfy ideal mixing ($\ket{\phi}=-\ket{s\bar{s}}$ and
$\ket{\omega}=\ket{u\bar{u}+d\bar{d}}/\sqrt{2}$) and (ii) the valence
quarks determine the low-$t$ behavior of the matrix elements
$\bra{0} J_\mu\ket{V}$ one expects
$\bra{0} \bar{s}\gamma_\mu s\ket{\omega}=0$ and
$\bra{0} \bar{s}\gamma_\mu s\ket{\phi}= -3 \bra{0} J_\mu^{EM}\ket{\phi}$.
It is straightforward to account for deviations from ideal mixing
\cite{Jaf89,MuI97,HaM98}:
\begin{eqnarray}
\label{kff:res}
C_\omega^{(s)}/C_\omega &=& -\sqrt{6}\left[{\sin\epsilon\over\sin(\epsilon+
\theta_0)}\right]\approx -0.2\,, \\
C_\phi^{(s)}/C_\phi &=& -\sqrt{6}\left[{\cos\epsilon\over\cos(\epsilon+
\theta_0)}\right]\approx -3\,, \nonumber
\end{eqnarray}
where the $s$ superscript denotes the residue for the strangeness form
factor, $\theta_0$ is the \lq\lq magic" octet-singlet mixing angle giving
rise to pure $u\bar{u}+d\bar{d}$ and $s\bar{s}$ states and $\epsilon$ 
deviations from ideal mixing. From Eqs. (\ref{kff:vmd}) and (\ref{kff:res}) 
we observe that the time-like kaon strangeness form factor is dominated by 
the $\phi(1020)$ resonance. We note that the flavor rotation of
Eq. (\ref{kff:res}) only gives the relative size of the $\omega$
and $\phi$ contributions  but does not lead to the correct normalization 
for $\fks$ at $t=0$ which must be enforced by hand.

\begin{figure}[htb]
\epsfxsize=4.5in
\begin{center}
\ \epsffile{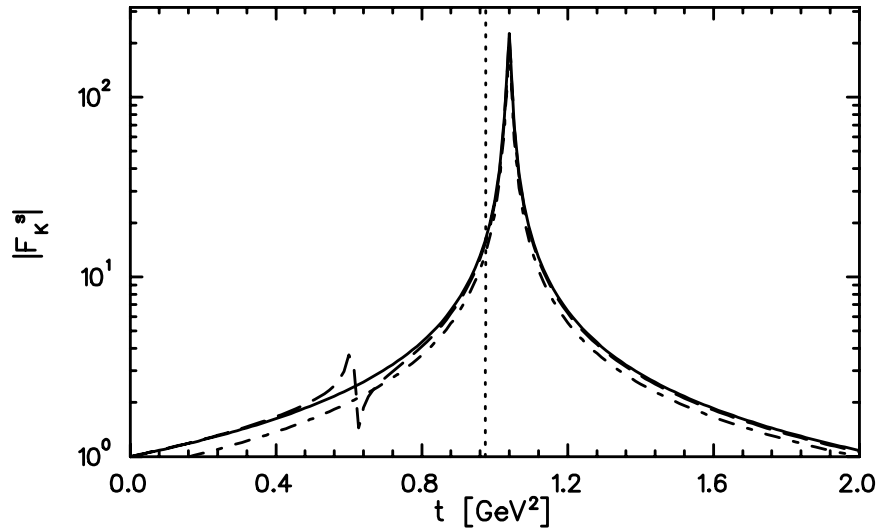}
\end{center}
\caption{\label{drsd:fksfig}Different model parametrizations for $\fks$.
Full line shows simple VMD model with the $\phi$ only, dashed
line shows the flavor rotated VMD model including the $\omega$ and
$\phi$, and dash-dotted line shows GS parametrization.
Dotted vertical line indicates $K\bar{K}$ threshold.}
\end{figure}

In Fig. \ref{drsd:fksfig} we plot $\fks$ as given by Eqs. (\ref{kff:vmd})
and (\ref{kff:res}) and compare it with a simple VMD model \cite{MHD97}
and the Gounaris-Sakurai (GS) parametrization for $F_\pi$ \cite{GoS68}
with all $\rho$ parameters replaced by the corresponding ones 
for the $\phi$.
We observe that the GS and VMD forms reproduce the essential features of
$\fks$ as determined from $e^+e^-$ data and standard flavor rotation 
arguments. Since $\fks$ is needed for $t\geq 4\mks$, the 
$\omega$-contribution which gives rise to the bump around 
$t\approx 0.6 \mbox{ GeV}^2$ in Fig. \ref{drsd:fksfig} is negligible.
In comparison to the strong $\phi$ peak, the small scale oscillations 
at higher-$t$ have a negligible impact as well. When computing 
the leading strangeness moments, we find less a than 10\% variation
in the results when any of these different parametrizations for $\fks$ is
used. In short, any model parametrization of $\fks$ showing the
peak at the $\phi$ mass and having the correct normalization,
$\fks(0)=-1$, may be used for the purpose of studying
nucleon strangeness. The pointlike approximation, $\fks = -1$,
however, misses important resonance physics. Throughout the
remainder of this work, we will use the GS parametrization.

\section{Analytic Continuation}
\label{sec:anacon}
To obtain the $\bll$  for $4\mks\leq t\leq 4\mns$ we
analytically continue physical amplitudes into the unphysical regime. 
The analytic continuation (AC) of a finite set of experimental amplitudes
with non-zero error is fraught with potential ambiguities. Indeed, 
AC in this case is inherently unstable, and analyticity alone has no
predictive power. Additional information must be used in
order to stabilize the problem, as we discuss below \cite{CPS75}.

In order to illustrate these issues and the methods we adopt to
resolve them, we first  briefly review the kinematics of $KN$ scattering
(cf. Ref. \cite{HWH97}). 
It is useful to consider the $s$-, $u$-, and $t$-channel
reactions simultaneously,\footnote{We do not consider $K^0 N$ scattering 
data in this analysis. In the following, we write $K$ and $\bar{K}$ for
$K^+$ and $K^-$, respectively.}
\beqa
\label{kkbar:chan}
&\mbox{(a) $s$-channel:} & \qquad K (q_i) + N(p_i) \rightarrow K 
(q_f) +N(p_f) \,,\\
&\mbox{(b) $u$-channel:} & \qquad \bar{K} (-q_f) + N(p_i) \rightarrow
\bar{K} (-q_i) + N(p_f) \,,\nonumber \\
&\mbox{(c) $t$-channel:} & \qquad \bar{K} (-q_f) + K (q_i) \rightarrow
\bar{N}(-p_i) + N(p_f) \,,\nonumber
\eeqa
where the four-momenta of the particles are given in parentheses.
In this notation the crossing relations between the different channels 
are immediately  transparent. The three processes can be described in 
terms of the usual Mandelstam variables
\beqa
s &=& (p_i+q_i)^2\,, \qquad
u = (p_i-q_f)^2 \,, \qquad\mbox{and}\qquad
t = (q_i-q_f)^2 \,.
\eeqa
The invariant matrix element for the $KN$ scattering process has
the structure,
\beq
\label{kkbar:defab1}
{\cal M}=\bar{u}(p_f)\left[ A(s,t)+\frac{1}{2}(\dida{q}_i+\dida{q}_f)
B(s,t)\right] u(p_i)\,,
\eeq
and the isospin decomposition of the invariant amplitudes $A$ and $B$ reads
\beqa
\label{kkbar:defab2}
A(s,t)&=&A^+(s,t) + 
A^-(s,t)\left(\vec{\tau}_N \cdot \vec{\tau}_K \right)\,,\\
B(s,t)&=&B^+(s,t) + B^-(s,t)\left(\vec{\tau}_N \cdot \vec{\tau}_K \right)
\nonumber \,.
\eeqa
We also need the $\Lambda$ and $\Sigma$ pole contributions to $A^{(\pm)}$ 
and $B^{(\pm)}$ which are given by
\beqa
\label{kkbar:polesu}
A^{(\pm)}_{\rm pole}(s,u)&=&\sum_{Y=\Lambda,\Sigma}\frac{g_\sst{KNY}^2}{2}
(m_\sst{Y}-\mn) \left(\frac{1}{u-m_\sst{Y}^2}\pm\frac{1}{s-m_\sst{Y}^2}
\right)\,,\\
B^{(\pm)}_{\rm pole}(s,u)&=&\sum_{Y=\Lambda,\Sigma}\frac{g_\sst{KNY}^2}{2}
\left(\frac{1}{u-m_\sst{Y}^2}\mp\frac{1}{s-m_\sst{Y}^2}\right)\,.\nonumber
\eeqa
\begin{figure}[htb]
\epsfxsize=3.5in
\begin{center}
\ \epsffile{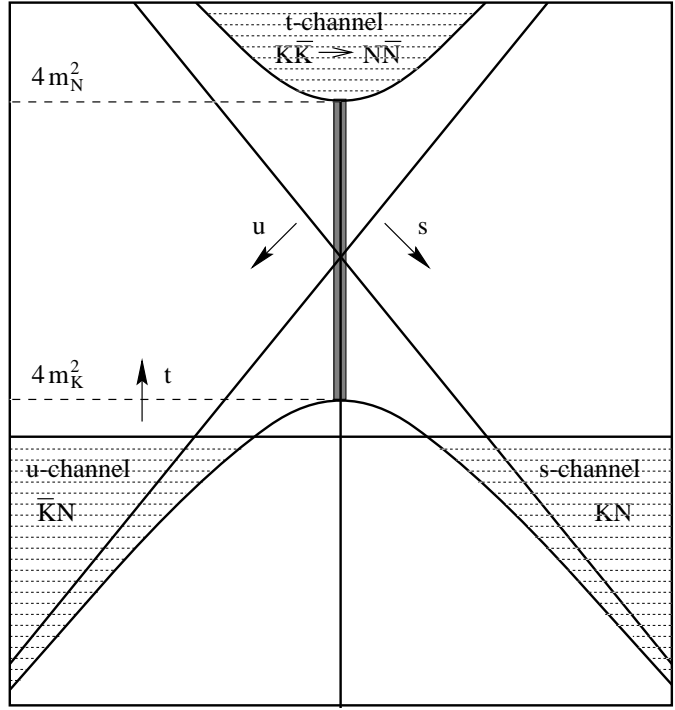}
\end{center}
\caption{\label{kkbar:fig1}Mandelstam plane for $KN$ scattering.
Physical regions are marked by dashed areas.}
\end{figure}

It is instructive to display the range of $s,u,$ and $t$ in the
Mandelstam plane, shown in Fig \ref{kkbar:fig1}. The physical
regions of the three reactions do not overlap, and
the invariant amplitudes simultaneously describe all three processes.
The physical values of the invariant amplitudes are obtained
when the Mandelstam variables are taken in the corresponding ranges.
In order to carry out the dispersion integrals of Eqs.
(\ref{drsd:fts}, \ref{drsd:fos}), we require the 
$\bll$ along the $t$-channel cut, 
indicated by the gray shaded area in Fig. \ref{kkbar:fig1}. 

We begin with experimental
$KN$ amplitudes in the s-channel region and employ the method of
backward DR to obtain the unphysical amplitudes
along the $t$-channel cut. The backward DR method has 
has been used successfully for a similar continuation
of $\pi N$ scattering amplitudes \cite{NLP70} and
as a consistency test for different $KN$ phase shift
solutions in the 1970's \cite{NiO72}. Only recently has a continuation
of the $KN$ amplitudes become possible due to improvements in the
data base over the last three decades. A
continuation of $KN$ scattering amplitudes
has also been performed on the basis of hyperbolic DR \cite{BrM79}.
In the latter analysis, however, the $t$-channel helicity
amplitudes have been parametrized by sharp resonance poles,
an a priori assumption we seek to avoid in the present analysis.
We use the recent $KN$ phase shift analysis of the
VPI-group \cite{Hys92} as experimental input.

\subsection{Overall Strategy and Problems}
Although an AC is uniquely defined from a
continuum of points, this is not the case for a finite set
of points lying within a certain error corridor.
Consequently, the procedure for obtaining amplitudes outside
the range of the given data is unstable. In practice the problem 
is stabilized by restricting the admissible solutions.
This is achieved by using a priori information on the solution
from experiment or theory as \lq\lq stabilizing levers" \cite{CPS75}.
In the remainder of this subsection, we give an overview of our
general procedure for making this stabilized continuation \cite{HWH97,RMH98}. 

The first important observation is that in the backward direction,
the invariant amplitudes in the $s$- and $t$-channel coincide, i.e.
$f_s(t,\theta_s=\pi)=f_t(t,\theta_t=\pi)=f(t)$.
Here $f$ is a generic
invariant amplitude and $\theta_s$ and $\theta_t$ are the CM-scattering
angles in the $s$- and $t$-channel, respectively. Thus, the continued
amplitude $f_s(t, \theta_s=\pi)$ in the unphysical regime can be identified
with the unphysical $t$-channel amplitude, $f_t(t,\theta_t=\pi)$.

Second, we work on a Riemann
sheet where only the singularities from the $s$- and
$t$-channel reactions are present. The backward
amplitude $f$ then possesses a left-hand cut from zero to minus infinity
stemming from the $s$-channel reaction. The lowest
right-hand cut, which is due to the $3\pi$ intermediate state in the
$t$-channel, runs from $9m_{\pi}^2$ to infinity.
The two cuts do not overlap and we can write an unsubtracted DR for $f$.
The amplitude is known from experiment on
the left-hand cut in the region $t_p \leq t \leq 0$ and we need its 
values on the right-hand cut where $f$ is
related to the $K\bar{K}\to N\bar{N}$ partial waves.
In particular, we are interested in the unphysical region $4\mks
\leq t \leq 4\mns$ where the partial waves are not bounded by unitarity.

Following our strategy to include as much a priori information as
possible, we do not continue the amplitude $f$ itself but rather
a related function, $\Delta f$, the so-called discrepancy function. 
In $\Delta f$ the analytically determined pole terms for $f$, as well
as the experimentally known behavior of the amplitude, have been subtracted
out. This subtraction is carried out in such a way as to remove the portion 
of the left-hand cut lying in the range $t_p\leq t\leq 0$ 
where the phase shift analyses are available. Thus, the use of 
$\Delta f$ presents two advantages. First, the original problem which 
entailed an AC from the boundary of the region of analyticity to another 
point on the boundary ($B\to B$) has been transformed into one in which 
the continuation occurs from the interior of the analyticity domain to the 
boundary ($I\to B$). Second, the well-known pole terms 
can be continued explicitly, and only non-pole parts which remain
in $\Delta f$ must be continued with other methods. 
In the end, $f$ is easily reconstructed from $\Delta f$.

The continuation of $\Delta f$ is carried out by means of a power
series expansion. Such an expansion, however, converges only up to
the nearest singularity. In our case this point lies at the
three-pion cut, well below the region of interest. We circumvent this
problem by using a conformal mapping of the complex $t$-plane. First
we symmetrize the cuts to lie along $(-\infty,-R]$ and $[R,\infty)$.
We then map the cuts for $\Delta f$ onto the circumference of the unit 
circle in the $w$-plane and the rest of the $t$-plane into its interior 
using 
\beq
t(w)=\frac{2Rw}{1+w^2}.
\eeq
Now it is possible to use a Legendre series in $w$ to perform the 
continuation.
In principle, the series can be fit to $\Delta f$ in the region 
$t_p \leq t \leq 0$ and the fit then be evaluated in the region of 
interest. In practice, we fit to the function  
\beq
\Delta g(w) = |t|^{-\alpha} \Delta f(t(w))
\eeq
instead of $\Delta f$. In so doing, we explicitly enforce the 
high-energy behavior of the amplitude $f$: $f(t)\sim |t|^{\alpha}$ for 
$t\to\infty$. The parameter $\alpha$ may be obtained from Regge
analyses. In addition, the explicit inclusion of the high-energy behavior 
stabilizes the continuation by suppressing oscillations in the solution. 
In fact, $\alpha$ can also be thought of as an auxiliary 
parameter introduced just for this purpose. The final results
depend only weakly on the precise value of $\alpha$. In fact, we evaluate
the stability of the continuation by studying the variation of the results
with $\alpha$ as well as with the number of terms included in the Legendre
series. The technical details of the AC procedure are given in 
Ref. \cite{HWH97}. In the next subsection, we directly start
with the continued backward amplitudes.

\subsection{Results}
\label{sec:kkbar:res}
First we introduce the invariant amplitude $F^{(\pm)}$ which
in the backward limit is related to $A^{(\pm)}$ and $B^{(\pm)}$
via
\beq
\label{kkbar:deff}
F^{(\pm)}(t) = A^{(\pm)}(t)+\mn\sqrt{q_t/p_t}\,B^{(\pm)}(t)\,.
\eeq
In order to extract the $\bll$, we need $B^{(+)}$ and $F^{(+)}$ 
in the $t$-channel unphysical region.
\begin{figure}[htb]
\epsfxsize=15cm

\vspace{-0.5cm}

\begin{center}
\ \epsffile{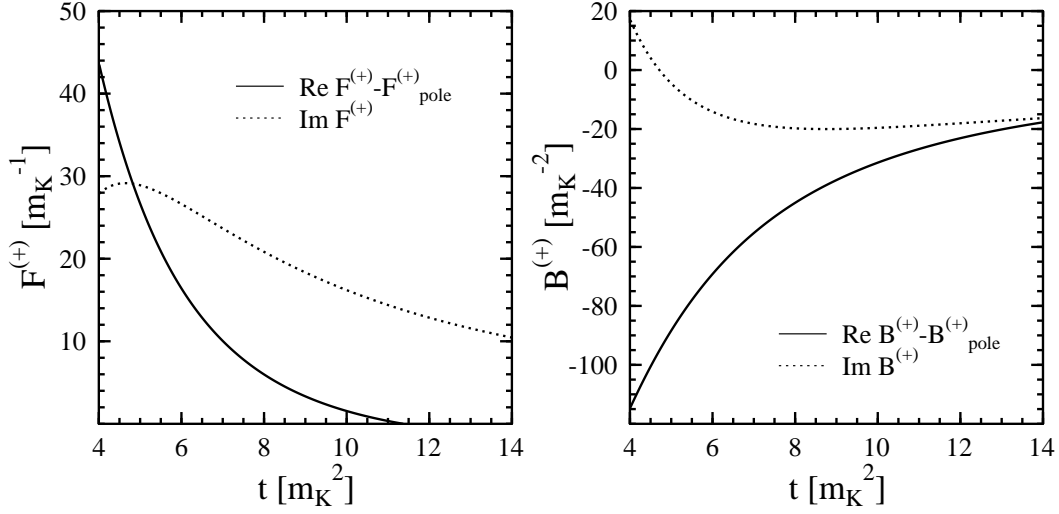}
\end{center}

\vspace{-1cm}

\caption{\label{kkbar:fig6}Analytic continuation of the invariant
backward amplitudes $\tilde{F}^{(+)}=F^{(+)}-F^{(+)}_{pole}$ (with
$n=6$) and $\tilde{B}^{(+)}=B^{(+)}-B^{(+)}_{pole}$ (with $n=5$)
for $\alpha=-1.2$.}
\end{figure}
In Fig. \ref{kkbar:fig6}, we show the results of the analytic
continuation for $\tilde{F}^{(+)}=F^{(+)}-F^{(+)}_{pole}$
and $\tilde{B}^{(+)}=B^{(+)}-B^{(+)}_{pole}$ up to $t=14\,\mks$.
The continued amplitudes vanish as $t \to \infty$.
We do not expect the power series expansion to be credible in a
larger region than the one in which its coefficients have been
determined. Since the experimental amplitudes
are given in the interval $-8\,\mks\leq t\leq 0$, we trust our
continuation only up to $t \approx 8\,\mks$ which covers half
of the unphysical region. Fortunately, for purposes of analyzing
nucleon form factors, the $K\bar{K}$ threshold region
dominates. The kaon form factor $\fka(t)$, which multiplies the
$\bll$ in the spectral functions 
amplifies the low-$t$ contribution and suppresses that from the
region $8\mks\leq t\leq 4\mns$. We plot only the results for the
amplitudes without their pole part. Since the pole part is exactly
known, it can be continued explicitly and is added to the 
remainder obtained from the AC in the end.

Ultimately, we require the $J=1$ partial wave projections
of the backward amplitudes which can be expanded in partial waves as
\beqa
\label{kkbar:pwb}
B^{(+)}(t)&=&\frac{8\pi}{q_t^2}\sum_{J=1}^\infty
\frac{J+\frac{1}{2}}{\sqrt{J(J+1)}} {b_J^{1/2,\,-1/2}}(t) P_J'(-1)\,,\\
F^{(+)}(t)&=&-\frac{4\pi\sqrt{t}}{p_t q_t}\sum_{J=0}^\infty
(J+\frac{1}{2}) {b_J^{1/2,\,1/2}}(t) P_J(-1)\,. \nonumber
\eeqa
Because the AC gives the invariant amplitudes only in 
the backward direction, it is non-trivial to extract the partial waves.
The expansion can be carried out separately for $\tilde f$ and 
$f_{\rm pole}$. This is useful because
the sum in Eq. (\ref{kkbar:pwb}) converges much faster for the
non-pole part than for the pole part \cite{NLP70}. Consequently,
we exploit the faster convergence for the non-pole part and add the
exactly known pole term projections after the $\bll$ have
been isolated.

Since the amplitude is known only for one value of $\theta_{t}$,
a separation of the $\bppm$ from the $b_{J\not= 1}^{\lambda,\bar{\lambda}}$
relies on several additional observations. First, 
each $|b_J^{\lambda,\bar{\lambda}}| \to 1$ or smaller as $t\to 4\mns$ 
because of unitarity. The only significant deviation
from this trend for $4\mks\leq t\leq 4\mns$ occurs via resonant enhancements
of the partial waves. The lightest $J>1$, $I=0$ resonance having a
non-negligible branching ratio to the $K\bar{K}$ state is the $f_2(1270)$, 
whose mass lies near the upper end of the range of validity of the AC.
Moreover, in Ref. \cite{BrM79} no evidence for $J\geq 2$ resonance 
effects close to the $K\bar{K}$ threshold 
was found. Consequently, we truncate the expansions 
in Eq. (\ref{kkbar:pwb}) at $J=1$. In order to test the validity of this 
truncation, we also perform an analysis with a model for the resonant $J=2$ 
partial waves included. As discussed below, our results are essentially 
unaffected by this inclusion. Since the width of the $f_2(1270)$ 
is $\sim 185$ MeV, we may expect some small, residual 
contamination of the $J\leq 1$ partial waves due to our truncation.
Fortunately, the presence of $\fka(t)$ in Eqs. 
(\ref{drsd:imf1}-\ref{drsd:imgm}) protects the spectral functions.
The kaon form factors are peaked in the vicinity of the $\phi(1020)$ and 
strongly suppress contributions from  $t > m_\phi^2$ \cite{MHD97}.

The remaining separation between the S- and P-waves in $B^{(+)}$ may be
performed by drawing on the work of Refs.
\cite{NiO72} and \cite{BrM79}, where backward and hyperbolic DR
have been used to analyze the $K\bar{K}\to N\bar{N}$ amplitudes
under the assumption that the helicity amplitudes are dominated
by sharp resonance poles. In the S-wave, one finds a resonance close 
to the $K\bar{K}$ threshold having a 22\% branching ratio to $K\bar{K}$, 
namely the $f_0(980)$. Therefore, we use the resonant
$b_0^{1/2,1/2}$ amplitude from Refs. \cite{NiO72,BrM79} and
subtract it from our result.\footnote{We can not use
the results of Refs. \protect\cite{NiO72,BrM79} directly because the
$\bll$ were assumed to be dominated by a single effective $\omega$ pole
and the explicit effect of the $\phi$ was {a priori} excluded.}
Both analyses use the following form for the amplitude
\beq
\label{kkbar:hmod}
b_0^{1/2,1/2}(t) = \frac{2 q_t}{\sqrt{t} p_t}
\frac{\Gamma_+^{f_0}}{m_{f_0}^2 -t -i\epsilon}\,,
\eeq
and obtain approximately the same result for the residue $\Gamma_+^{f_0}$.
We use $\Gamma_+^{f_0} = (-24.10 \pm 6.65 )\, m_\sst{K}^3$ \cite{BrM79}
and extract the $\bll$ from the continued invariant amplitudes using
\beqa
\label{kkbar:extra}
\bpm(t) &=&\frac{\sqrt{2} q_t^2}{12\pi} \tilde B^{(+)}(t)+
\left.\bpm(t)\right|_{\rm pole}\,,\\ \bpp(t) 
&=&\frac{1}{3}\left( \frac{i p_- q_t}{2\pi\sqrt{t}} \tilde F^{(+)}(t)
+ b_0^{1/2,1/2}\right)+\left.\bpp(t)\right|_{\rm pole}\,,\nonumber
\eeqa
with $p_- = \sqrt{\mns-t/4}$. Note that the $\Lambda$ and $\Sigma$ pole
term projections for even values of $J$ vanish for the isoscalar amplitudes 
in the $t$-channel. As a consequence, the pole term projection for $J=0$ 
does not appear in Eqs. (\ref{kkbar:extra}). The $J=1$ pole
term projections are
\beqa
\label{kkbar:ptp}
\left.\bpm(t)\right|_{\rm pole}&=&\sum_{Y=\Lambda,\Sigma}\frac{\sqrt{2}
q_t g_\sst{KNY}^2}{24 \pi i p_-}\left(Q_0(-i \xi_\sst{Y})-Q_2(
-i \xi_\sst{Y})\right) \,,\\
\left.\bpp(t)\right|_{\rm pole}&=&\sum_{Y=\Lambda,\Sigma}
\frac{g_\sst{KNY}^2}{ 4 \pi\sqrt{t}}\bigg[ (m_\sst{Y}-\mn)\,
Q_1(-i \xi_\sst{Y}) \nonumber\\& &
\hphantom{\sum_{Y=\Lambda,\Sigma}}
+\frac{\mn q_t}{3 i p_-}\left(2 Q_2(-i \xi_\sst{Y})+Q_0(-i \xi_\sst{Y})
\right) \bigg] \,, \nonumber
\eeqa
where the $Q_i$, $i=0,1,2$ are Legendre functions of the second kind and
$\xi_\sst{Y}$ is given by
\beq
\label{kkbar:ptpxi}
\xi_\sst{Y}=\frac{t-2\mks+2(m_\sst{Y}^2-\mns)}{4 q_t p_-}\,.
\eeq
The introduction
\begin{figure}[htb]
\epsfxsize=15.5cm
\begin{center}
\ \epsffile{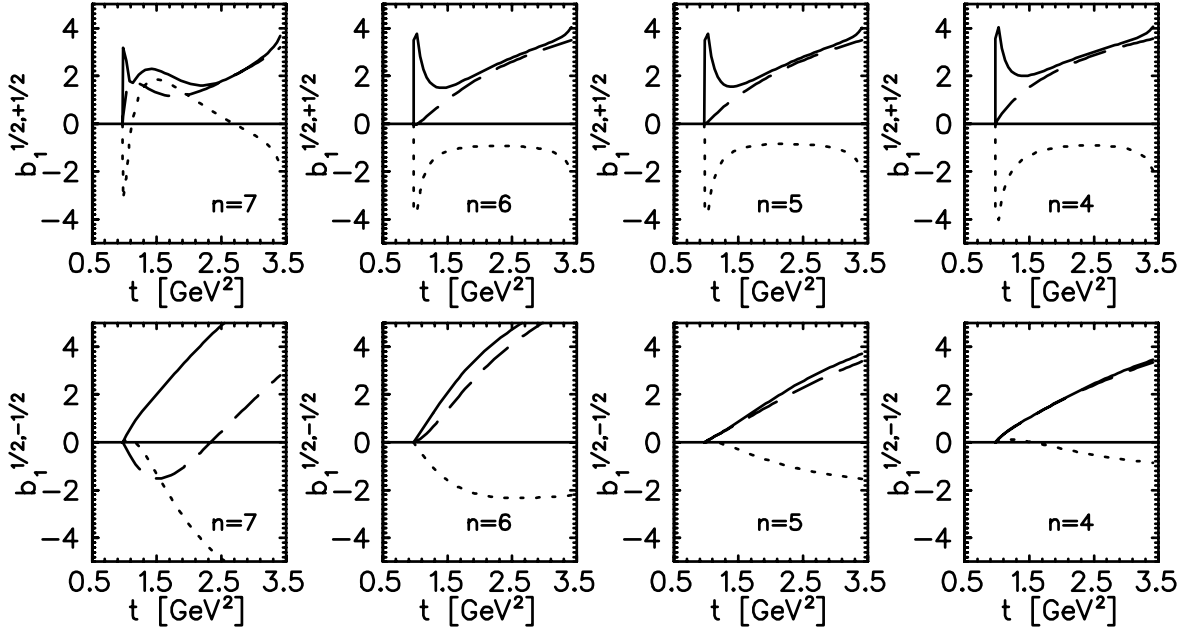}\
\end{center}
\caption{\label{kkbar:fig7}Comparison of the results for the
$b_1^{1/2,\,1/2}$ (upper row) and $b_1^{1/2,\,-1/2}$ (lower row)
for different order of the Legendre series $n$ and $\alpha=-1.2$.
Solid lines show $|b_1^{1/2,\,\pm 1/2}|$,
dashed lines show $\Re\,b_1^{1/2,\,\pm 1/2}$, and
dotted lines show $\Im\,b_1^{1/2,\,\pm 1/2}$.The continuation is 
expected to be trustworthy up to $t\approx 2\mbox{ GeV}^2$.}
\end{figure}
of a width in Eq. (\ref{kkbar:hmod}) changes the
extracted amplitudes $b_1^{1/2,\,\pm1/2}$ only slightly and has almost
no impact on the application to nucleon form factors. The same
observation applies for the case when the $J=2$ partial waves from
Ref. \cite{BrM79} are subtracted as well. Therefore, we discard the 
$J\geq 2$ partial waves and proceed accordingly. 

In the following, we discuss the sensitivity of the
amplitudes to the order $n$ of the Legendre series and
the high-energy parameter $\alpha$. In Fig. \ref{kkbar:fig7},
we display the full $b_1^{1/2,\,\pm1/2}$ for different values of $n$.
We show their real and imaginary parts as well as their absolute value.
It is clearly seen that $b_1^{1/2,\,1/2}$ is almost independent of
$n$, whereas $b_1^{1/2,\,-1/2}$ shows some variation.
From the quality of the fits to the discrepancy functions,
we choose $n=6$ for $F^{(+)}$, whereas we take $n=5$ for $B^{(+)}$.
This choice corresponds to taking the minimum $n$ which
gives a satisfactory fit in order to minimize the amplification
of experimental noise \cite{CPS75,Sab80}.
The dependence of our results on the asymptotic parameter
$\alpha$ is similar. Although $\alpha$ is a physical parameter, 
its determination is model dependent.
We have varied $\alpha$ from $-0.2$ to $-5.0$ to test the 
sensitivity of the AC to $\alpha$. Similar to the dependence 
on $n$, the $b_1^{1/2,\,1/2}$ are almost independent of $\alpha$, 
whereas the $b_1^{1/2,\,-1/2}$ show some variation.
For $\alpha \to 0$ our analysis becomes unstable as we expect
since arbitrary oscillations in the AC are no longer 
suppressed. Furthermore, for $\alpha \geq 0$ the DR
does not converge and the corresponding results are meaningless. 
For the final results we take $\alpha=-1.2$ from the Regge-model fit of
Ref. \cite{Bar69}. From the dependence on $n$ and $\alpha$,
we expect our results for $b_1^{1/2,\,1/2}$ to be more reliable.
This conclusion is supported by the fact that 
$|F|^2$ is directly related to the backward cross section.
Hence, $F$ is particularly well determined by the data
and errors in the phase shift analysis cancel in the reconstruction
of $F$ \cite{NLP70,NiO72}. 

As shown in Fig. \ref{kkbar:fig7}, a
clear resonance structure at threshold is seen in $b_1^{1/2,\,1/2}$, which
presumably is the $\phi$ resonance. This resonance is not observed
in $b_1^{1/2,\,-1/2}$. There are two possible explanations for this 
fact:

\medskip
\noindent 1. The AC for $B$ and consequently $b_1^{1/2,\,-1/2}$
is not sufficiently well determined for the reasons explained above.

\medskip
\noindent 2. The vector and tensor couplings of the $\phi$ meson to
the nucleon are approximately equal and have opposite signs.
Parametrizing the resonant part of the $J=1$ partial waves
by the $\phi$, we have \cite{BrM79}
\beqa
\label{kkbar:phired}
\Gamma^\phi_- &=& -\frac{2\sqrt{2}}{3}\frac{G_{\phi K\bar{K}}}{4\pi}
\left(G^V_{\phi N\bar{N}}+G^T_{\phi N\bar{N}}\right)\,,\\
\Gamma^\phi_+ &=& -\frac{2\mn}{3}\frac{G_{\phi K\bar{K}}}{4\pi}
\left(G^V_{\phi N\bar{N}}+\frac{m_\phi^2}{4\mns}G^T_{\phi N\bar{N}}
\right)\, ,\nonumber
\eeqa
where the $\Gamma^\phi_{\pm}$ are the residues corresponding
to $b_1^{1/2,\,\pm 1/2}$ (cf. Eq. (\ref{kkbar:hmod})). 
$G^V_{\phi N\bar{N}}=-G^T_{\phi N\bar{N}}$ then leads to a 
resonance in $b_1^{1/2,\,1/2}$ but not in $b_1^{1/2,\,-1/2}$.
\medskip

\noindent
Following scenario 2 above, we determine $G_{\phi N\bar{N}}^V$ and $G_{\phi
N\bar{N}}^T$ from the resonance structure in $\bpp$. We fit the region 
around this structure  using the $J=1$ analog of Eq. (\ref{kkbar:hmod}), 
but including a finite width, $\Gamma_\phi$. 
We obtain $G_{\phi N\bar{N}}^V=7.4$ using the
Particle Data Group value for $\Gamma_\phi$ \cite{PDG98} and 
$G_{\phi N\bar{N}}^V=9.6$ when $\Gamma_\phi$ is allowed to be a fit 
parameter. These values are comparable with the value $G_{\phi N\bar{N}}^V
=9.2$ obtained from the VMD parametrization of
$\FOis$ \cite{MMD96}. Our value for $G_{\phi N\bar{N}}^T=-
G_{\phi N\bar{N}}^V$, however, has a larger magnitude than
obtained in Ref. \cite{MMD96}. The possible reasons for this
difference are discussed elsewhere \cite{RMH98x}. We
emphasize, however, that our determination of the $\phi N\bar{N}$ couplings
relies only on the observed resonance structure in the strong amplitude
$\bpp$ and not on a VMD ansatz for the isoscalar EM form factors.
We also note that the {\em interpretation} of the $\bll$ in terms of
$\phi N\bar{N}$ couplings is inconsequential for the computation of the 
$F_i^{(a)}$, since the amplitudes themselves -- rather than a
parametrization of them -- are used in the dispersion integrals.

\section{$K\bar{K}$ contribution to Nucleon Form Factors}
\label{sec:kkbarnucff}
We now use the $\bll$ to evaluate the contribution of the
unphysical region to the dispersion integral for the strange
and isoscalar nucleon form factors. 
We consider the electric and magnetic Sachs radii and the
anomalous magnetic moment. Drawing upon Eqs. (\ref{drsd:fts},
\ref{drsd:fos}), we write down
subtracted DR's for the Sachs radii,
\beq
\label{kkbar:rsstrdr}
{\rho^a_i} =-{4\mns\over\pi}\int_{4\mks}^\infty\ dt'\,
            {\Im\, G_i^a(t')\over t^{2}}\, ,
\eeq
where $i=E,M$ and $a$ denotes the isoscalar EM or strangeness channel. 
Before evaluating
Eq. (\ref{kkbar:rsstrdr}), we discuss the qualitative features of
the spectral functions for $\rho^a_E$, $\rho^a_M$, and $\kappa^a$.
Since for our purpose the difference between the EM and 
strange kaon form factors is essentially given by the normalization, 
the qualitative features of the strange
and EM spectral functions are the same. Consequently,
we limit the following discussion to the strange spectral functions.

\begin{figure}[ht]
\epsfxsize=14cm
\begin{center}
\ \epsffile{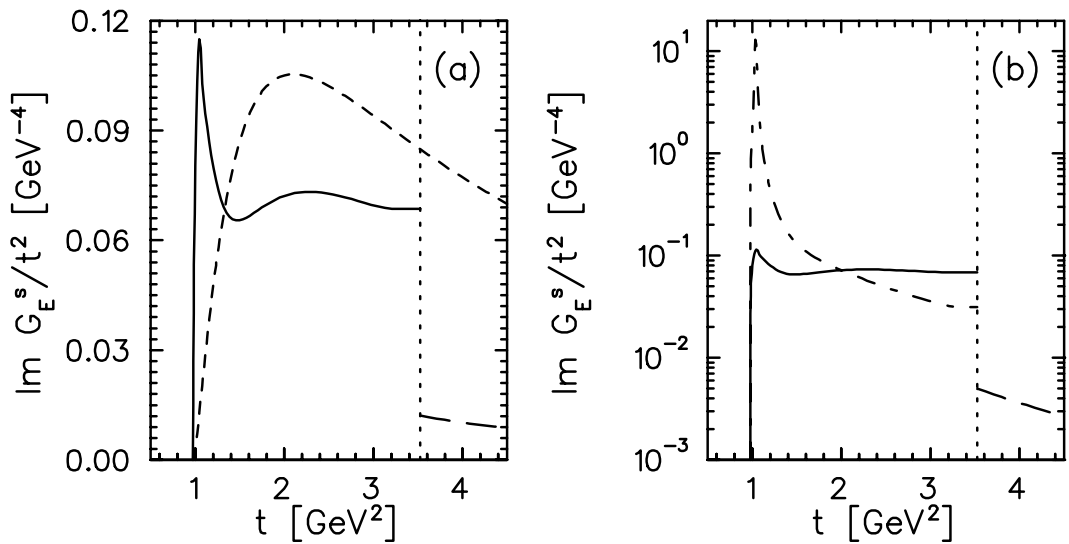}
\end{center}
\caption{\label{kkbar:sprhose}$K\bar{K}$ contributions to the 
spectral function for $\rho_E^s$.
Short dashed curve (a) gives ${\cal O}(g^2)$ result of scenario
(I) discussed in the text. Solid curve [(a) and (b)] gives result for
scenario (II). Long-dashed curve gives unitarity bound for $t\geq 4\mns$
using a pointlike $\fks$ (scenario (II)) and GS parametrized $\fks$
(scenario (III)). Dashed-dotted curve (b) gives all-orders spectral
function (scenario (III)). Dotted vertical line indicates physical
$N\bar{N}$ production threshold.
}
\end{figure}

The $K\bar{K}$ content of the spectral function for
$\rho_E^s$ is displayed in Fig. \ref{kkbar:sprhose}.
In Fig. \ref{kkbar:sprhose}a, we compare two scenarios for $\bpp$:
(I) the Born approximation in the nonlinear $\sigma$-model (BA)
and (II) the AC from the previous section.
For both scenarios a pointlike kaon form factor $\fka(t)\equiv 1$ has been
used. Although the pointlike form factor is unrealistic, using it in
this context  allows to illustrate separately the effects of form factor
and scattering amplitudes.

We include scenario (I) because of its correspondence with a number of
model calculations as well as CHPT. The scenario (I) spectral function
contains only contributions to ${\cal O}(g^2)$, where $g$ is the scale
of the strong hadronic couplings. It constitutes the DR form of a one-loop
calculation containing a kaon and strange baryon intermediate state and
a current insertion on the kaon line \cite{MHD97,FGT58}. A variety of model
calculations (see, e.g., Refs. \cite{Mus94,MuB94})
have been performed under the assumption that such amplitudes dominate the
strangeness form factors and that a truncation at ${\cal O}(g^2)$ gives a
reliable estimate of the scale and sign of the kaon contribution. 
In ChPT analyses of both the strange and isoscalar EM moments,
the non-analytic (in quark masses) parts of the same amplitudes are 
retained and added to the appropriate low-energy constants.
In the case of $\rho_E^s$, for example, the leading-order non-analytic 
contribution is singular in the chiral
limit \cite{MuI97}:
\begin{equation}
\label{eq:chiralsing}
\rho^s_{non-anal}=-\left({\mn\over\Lambda_\chi}\right)^2
\left\{1+{5\over 3}\left[\left({3F+D\over\sqrt{6}}\right)^2+
{3\over 2}(D-F)^2\right]\right\}\ln{m_K^2\over \mu^2}\, ,
\end{equation}
where $F$ and $D$ are the usual SU(3) reduced matrix elements,
$\Lambda_\chi=4\pi f_\pi$ gives the scale
of chiral symmetry breaking, and $\mu$ is
a renormalization scale. The chiral singularity of Eq. (\ref{eq:chiralsing})
arises from the kaon-nucleon one loop graph with the strange vector current
inserted on the kaon line. The equivalence of this loop calculation with the
scenario (I) for the DR analysis \cite{MHD97,FGT58,DrZ60} guarantees that the
DR computation includes $\rho^s_{non-anal}$.

At leading order ChPT, both $\rho_E^s$ and $\kappa^s$ also contain low-energy
constants which cannot be obtained from existing data using
chiral symmetry \cite{MuI97}. For the magnetic radius, however,
only the non-analytic {\em one-loop} terms contribute at
${\cal O}(p^3)$ \cite{HMS98}. In principle, the low-energy constants
include higher-order effects (in $g$) not contained in the non-analytic
loop contributions. A comparison of the scenario (I) spectral
function with the full spectral
function -- containing contributions to all orders in $g$ -- allows us to
determine the importance of these low-energy constants and more generally to
evaluate the credibility of one-loop predictions.

For both scenarios (I) and (II), we show the
upper limit on the spectral function generated by the unitarity bound
on $\bpp$.  As observed previously in Ref. \cite{MHD97},
and as illustrated in Fig. \ref{kkbar:sprhose}a,
the BA omits these rescattering corrections and consequently
violates the unitarity bound by a factor of four or more even at
the $N\bar{N}$ threshold. This feature alone casts a shadow on
predictions which rely {\em solely} on one-loop amplitudes. 
In principle, the low-energy constants of ChPT correct for the
unitarity violation implicit in the one-loop contributions. The
counterterm-free ${\cal O}(p^3)$ prediction for $\rho_M^s$, however,
does not include this unitarity correction.

The curve obtained in scenario (II) indicates
the presence of a peak in the vicinity of the $\phi(1020)$
meson, reflecting the presence of a $K\bar{K}\leftrightarrow
\phi$ resonance in $\bpp$.
This structure enhances the spectral function over the ${\cal O}(g^2)$ 
result near the $K\bar{K}$ threshold.
As $t$ increases from $4\mks$,
the spectral function obtained in scenario (II) falls below that
of scenario (I), presumably due to $K\bar{K}$ rescattering which must
eventually bring the spectral function below the unitarity bound for
$t\geq 4\mns$. The fact that the AC is approximately
constant above $t\simge 2 \mbox{ GeV}^2$ and also violates the
unitarity bound, indicates that the AC and partial wave
separation cannot be trusted close to the $N\bar{N}$ threshold. 

In Fig. \ref{kkbar:sprhose}b we plot the spectral function for
a third scenario (III), obtained by including a 
realistic (empirical) kaon form factor. We
use the GS parametrization, although other parametrizations
for $\fks$, such as a simple $\phi$-dominance form, yield similar
results as has been discussed above and elsewhere \cite{MHD97}.
The corresponding spectral function is shown in
Fig. \ref{kkbar:sprhose}b, and compared with the spectral 
function for a pointlike $\fks$. The use of the 
GS parametrization significantly enhances the spectral function near the 
beginning of the $K\bar{K}$ cut as compared with the pointlike case, while 
it suppresses the spectral
function for $t\simge 2\mbox{ GeV}^2$. Consequently, the full spectral
function is dominated by the low-$t$ region where the AC for 
the $\bll$ is reliable. The $t\simge 2\mbox{ GeV}^2$
region gives a negligible contribution, even though the 
$\bll$ are too large in this range (see Fig. 
\ref{kkbar:sprhose}b). The error associated with this region is 
correspondingly negligible. We emphasize that the one-loop
model predictions miss entirely the resonance enhancement of the 
$K\bar{K}$ contribution.   

\begin{figure}[ht]
\epsfxsize=14cm
\begin{center}
\ \epsffile{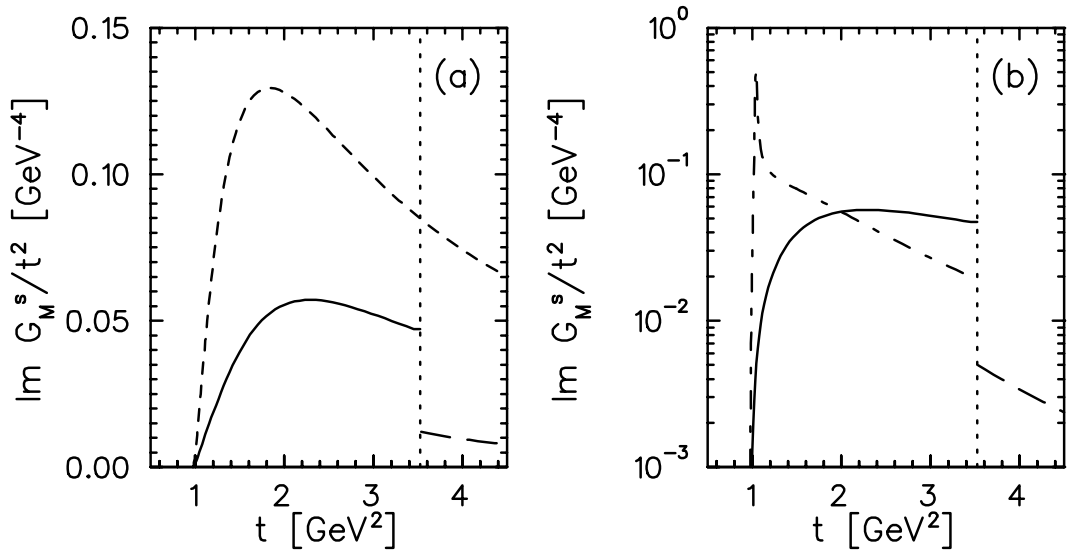}
\end{center}
\caption{\label{kkbar:sprhosm}$K\bar{K}$ contribution to spectral 
function for $\rho_M^s$. Curves are as in Fig. \ref{kkbar:sprhose}.}
\end{figure}

The corresponding scenarios for the magnetic Sachs radius $\rho_M^s$
are shown in Fig. \ref{kkbar:sprhosm}. In contrast to $\bpp$, $\bpm$
shows no resonance structure at all. The possible explanations for
this behavior have been discussed in the previous section. Here
we take our result at face value and develop the pertinent consequences
for $\rho_M^a$.
In Fig. \ref{kkbar:sprhosm}a we compare the results of scenario (I)
BA and (II) AC for a pointlike coupling of the kaon to $\bar{s}\gamma_\mu 
s$. The AC spectral function stays clearly  below the BA result
for all $t$, presumably due to rescattering effects included in the
AC. Similar to the electric case the unitarity
bound is also violated in scenario (II), although the violation is about
a factor two weaker than in the BA. Again we conclude
that the continued $\bpm$ is not trustworthy close to $t=4\mns$.
The suppression of the spectral function for $t\simge 2\mbox{ GeV}^2$
by the GS form factor in scenario (III) is displayed in Fig.
\ref{kkbar:sprhosm}b.

The absence of the resonance structure in $\bpm$ suggests that the 
integrated spectral function for scenario (III) gives an upper bound,
rather than a firm prediction, for ${\rho_M^a}$. The reason has to
do with the relative phases of $\bpm$ and $\fka(t)$. Consider
the strangeness case with a simple VMD parametrization of
$\fks(t)$ (other parametrizations are similar):
\begin{equation}
\fks(t) = {m_\phi^2\over m_\phi^2-t-im_\phi\Gamma_\phi}\, .
\end{equation}
As input to Eqs. (28-32), we require
\begin{equation}
\Re\, \bll\fks^{\ast} = (\Re\ \bll)(\Re\ \fks^{\ast}) +
        (\Im\ \bll)(\Im\ \fks^{\ast})\, ,
\end{equation}
where $\Re\ \bll$ and $\Im\ \bll$ are shown in Fig. \ref{kkbar:fig7}.
As $t$ crosses the resonance in the vicinity of $m_\phi^2$, 
$\Re\, \fks^{\ast}$ changes sign while $\Im\, \fks^{\ast}$ does not.
Instead, the latter reaches its peak value of magnitude 
$\sim m_\phi/\Gamma_\phi$.  From Fig. \ref{kkbar:fig7},
we observe that
neither $\Re\, \bll$ nor $\Im\, \bll$ undergoes a phase change
around $t=m_\phi^2$. Hence, when integrated across the resonance,
the contributions to the integral from $(\Re\, \bll)(\Re\, \fks^{\ast})$
change sign, leading to substantial cancellations. The contributions
from $(\Im\, \bll)(\Im\, \fks^{\ast})$, on the other hand, do not
change sign, and no cancellations occur. In the case
of $\bpp$, it is $\Im\, \bpp$ which contains the resonance structure.
Consequently, the integral for 
$\rho_E^s$ is dominated by the resonating imaginary parts of 
$\bpp$ and $\fks$, which remain in phase. The continuum contribution
is suppressed by the relative phase change of the real parts.

The situation for $\rho_M^s$ is different because
$\bpm$ displays no resonant behavior. Again, the 
contribution to the integral from $(\Re\, \bpm)(\Re\, \fks^{\ast})$ is
suppressed by cancellations from the phase change across
the resonance. $|(\Im\,\bpm)(\Im\,\fks^{\ast})|\ll |\bpm\,\fks^{\ast}|$,
however, and its precise magnitude is  sensitive
to the parameters $\alpha$ and $n$. Consequently, the integral of 
$\Re\, (\bpm\fks^{\ast})$ is rather uncertain. We are confident,
therefore, in quoting only an upper bound for $\rho_M^s$, obtained
by integrating $|\bpm\, \fks^{\ast}|$, which varies only gently with
$\alpha$ and $n$, and using
$\gamma_K=1$ (Eq. (\ref{uni:rel_p})).  As we note below, 
even this upper bound is
nearly twice as small as the result obtained from scenario (I). 

To determine the anomalous magnetic moment $\kappa^a$, we require an
unsubtracted DR and turn to $F_2^a$.
Eq.(\ref{drsd:fts}) reduces in the limit $t=0$ to
\begin{eqnarray}
\label{kkbar:musdr}
{\kappa^a} = F_2^a(0) &=&{1\over\pi}
\int_{4\mks}^\infty\ dt'\, {\Im\, F_2^a(t')\over t}\, .
\end{eqnarray}
with $\Im\,F_2^a$ given in Eq. (\ref{drsd:imf2}).
However, an additional comment is in order. $\Im \, F_2^a$
depends on both $\bll$ rather than on
one as for the radii. In order to guarantee a finite spectral
function at the $N\bar{N}$ threshold, the two $\bll$ must
fulfill the threshold relation $\bpm=\sqrt{2}\bpp$. Our
AC, however, is not reliable at $t=4\mns$ and does not obey this
relation. Therefore, we replace $\Im \, F_2^a(t)$ by $\Im \, F_2^a(t=
8\mks)$ for $8\mks \leq t\leq 4\mns$.
Doing so leads to an upper bound for the spectral function.
Essentially the same qualitative observations as for $\rho_E^a$ apply
to $\kappa^a$ because the spectral function is dominated by the resonance 
in $\bpp$ in both cases. Specifically, the contribution from
$t \geq 8 \mks$ is negligible. 

The numerical consequences of our analysis are indicated in
Table \ref{kkbar:tab1}, where we give results for the leading strangeness
moments.
\begin{table}
\begin{center}~
\begin{tabular}{|cc||c|c|c|}\hline
Scenario & Moment & $4\mks\leq t\leq4\mns$ & $4\mns\leq t$ & Total
\\\hline\hline
 & ${\rho_E^s}$ & $0.25$ & $0.10$ & $0.35$ \\
${\cal O}(g^2)$ & ${\rho_M^s}$ & $0.30$ & $0.05$ & $0.35$ \\
 & $\kappa^s$ & $-0.07$ & $-0.07$ & $-0.14$ \\\hline
 & ${\rho_E^s}$ & $0.98$ & $0.01$ & $0.99$\\
 AC/GS & ${\rho_M^s}$ & $0.17$ & $0.01$ & $0.18$\\
 & $\kappa^s$ & $-0.41$ & $-0.01$ & $-0.42$ \\\hline
\end{tabular}
\vspace{0.5cm}
\caption{\label{kkbar:tab1} Kaon cloud contribution for ${\rho_E^s}$,
${\rho_M^s}$, and $\kappa^s$ for the scenarios discussed in the text.
Second and third columns give contributions to the dispersion integral
from the unphysical and physical region, respectively. In
both scenarios, the unitarity bound on the $\bll$  is imposed for
$t\geq 4\mns$. To convert $\rho^s$ to $\langle r^2\rangle^s$, multiply
$\rho^s$ by -0.066 fm$^{2}$.}
\end{center}
\end{table}
The first three lines give the ${\cal O}(g^2)$ \lq\lq kaon cloud" 
prediction for ${\rho_E^s}$, ${\rho_M^s}$, and $\kappa^s$ (scenario I).
The last three lines give the results when the GS
form factor and analytically continued fit to $K N$ amplitudes for
$b_1^{1/2,\,\pm1/2}$ are used (AC/GS).
In all cases the unitarity bound on $b_1^{1/2,\,\pm1/2}$ is imposed for
$t\geq 4\mns$. The AC/GS results were obtained by extending $\bpp$ to
$4\mns$. Although the continuation can be trusted only for
$t\simle 8\mks$, and although the continued amplitudes exceed the
unitarity bound for $t\to 4\mns$, the GS form factor suppresses
contributions for $t\simge 8\mks$, rendering the overall contribution
from $8\mks\leq t\leq 4\mns$ negligible. The overall sign of the
product of $\fks^{\ast}$ and the $\bppm$ has been determined by
fits to EM form factor data \cite{RMH98x}. However,
the sign of ${\rho_E^s}$ is well determined from the phase of
$\bpp$ and $\fks^{\ast}$ at the $\phi$ peak.

From Table \ref{kkbar:tab1}, we observe that the use of a
realistic $K\bar{K}$ spectral function increases the
kaon cloud contribution to ${\rho_E^s}$ by roughly a factor of three as
compared to the ${\cal O}(g^2)$ calculation. Moreover, the result of the
all-orders computation approaches the scale at which the
proposed PV electron scattering experiments 
\cite{Mus94} are sensitive, whereas the 
${\cal O}(g^2)$ calculation (e.g., one-loop) gives a result which 
is too small to be seen. This observation depends critically on 
the presence of the resonance structure near $t=m_\phi^2$
in both $\bpp$ and $\fks$ (Fig. \ref{kkbar:sprhose}).
Its absence from either the scattering amplitude
or kaon form factor would lead to a significantly
smaller magnitude for ${\rho_E^s}$.

Because $\bpm$ does not display any resonance structure, 
no enhancement in $\rho^s_M$ is observed.
In fact the ${\cal O}(g^2)$ result is a factor of two larger than in the 
AC/GS case. We emphasize that the AC/GS result for $\rho_M^s$ constitutes 
an upper bound, given the phase uncertainty discussed above. 
It is noteworthy that this bound $|\rho^s_M|=0.18$
is not consistent with the range given by ChPT to order ${\cal O}(p^3)$,
$\rho^s_M=2.44\ldots 9.05$ \cite{HMS98}. Although both the DR and ChPT 
calculations of $\rho_M^s$ include only 
the $K\bar{K}$ contribution, a large discrepancy
exists between the two methods. We suspect that the problem lies
in the truncation at ${\cal O}(p^3)$ in ChPT, for the following reasons.
First, the ${\cal O}(g^2)$ spectral function contains 
no rescattering corrections required for consistency
with the unitarity bound.  When integrated as in Eq. (\ref{kkbar:rsstrdr}), 
this spectral function yields the value for $\rho_M^s$ given in Table 
\ref{kkbar:tab1}. Second, the ${\cal O}(p^3)$ calculation in ChPT includes 
only the non-analytic contributions to this integral.
The latter yield a result that is an order of magnitude 
larger than the result obtained for the full ${\cal O}(g^2)$ integral. 
Given that this is already a factor of two larger than the 
upper bound obtained from the all orders DR result, we suspect that 
the ${\cal O}(p^3)$ ChPT result omits crucial higher-order rescattering 
effects. Presumably, these effects are contained in the 
${\cal O}(p^4)$ Lagrangian, which contains a term of the form
\beq
{\cal L}_{(4)} = {b\over\Lambda_\chi^3} {\bar\psi}\sigma_{\mu\nu}
\psi \partial^2 F^{\mu\nu}\,, 
\eeq
where $\Lambda_\chi=4\pi f_\pi$.
In effect, the AC/GS result of Table \ref{kkbar:tab1} represents a
determination of the kaon rescattering contributions to $\bll$. 
Unfortunately, ChPT cannot make counterterm free predictions for 
the other moments \cite{MuI97} and the problem remains unresolved.

In the case of $\kappa^s$, we observe an increase by a factor of three
over the ${\cal O}(g^2)$ approximation. However, due to the presence of
both $\bpp$ and $\bpm$ in $\Im\,F_2$ and the phase-related uncertainty
associated with $\bpm$, we consider the all-orders AC/GS value for
$\kappa^s$ to be an upper bound. Given the size of the experimental
errors, this bound is not incompatible with the SAMPLE results
(see Eq. (\ref{intro:sample})).

It is interesting to note that only the normalization of $\GMS$ at
$q^2=0$ but not its $q^2$-dependence receives a resonance-enhanced
$K\bar{K}$ contribution. This feature can be understood by
observing that the resonance-enhanced partial wave $\bpp$ enters
$\Im\, F_1$ and $\Im\, F_2$ with the same coefficient but opposite sign
(cf. Eqs (\ref{drsd:dsacsff}, \ref{drsd:imf1}, \ref{drsd:imf2})).

\begin{figure}[ht]
\epsfxsize=12cm
\begin{center}
\ \epsffile{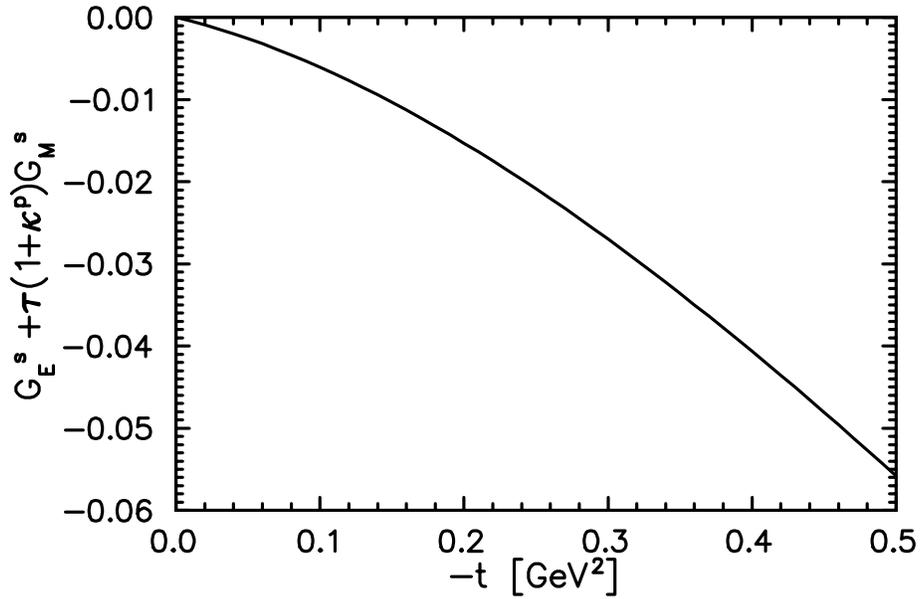}
\end{center}
\caption{\label{kkbar:sexp}
$K\bar{K}$ contribution to $\GES+\tau (1+\kappa^p) \GMS$ as function 
of the momentum transfer $t$. The contributions from $\GES$ and
$\GMS$ enter with opposite signs.}
\end{figure}

In Fig. \ref{kkbar:sexp},
we plot the quantity extracted from forward angle
parity-violating electron scattering experiments on the proton,
$\GES+\tau (1+\kappa^p)\GMS$,
as a function of the momentum transfer $t$. 
At $t=0.48\mbox{ GeV}^2$, one can compare the result of the 
HAPPEX collaboration (see Eq. (\ref{intro:happex})) with the
$K\bar{K}$ contribution:
$\left| \GES+0.39\GMS\right|_{K\bar{K}}=0.053$. 
The two values agree within the experimental error bars.\footnote{Note
that we quote only the absolute value because of the phase-related
uncertainty in $\GMS$.}

From a qualitative standpoint, the $K\bar{K}$ spectral content of
the isoscalar EM and strangeness form factors are similar. The
numerical significance, however, differs between the two cases.
In Table \ref{kkbar:tab2}, we give the $K\bar{K}$
contribution to the leading isoscalar EM moments. We compare the $K\bar{K}$
contributions for the AC/GS scenario with the total \lq\lq experimental'' 
values from the dispersion theoretical analysis of Ref. \cite{MMD96} and 
the $\phi$ pole contribution alone.\footnote{If a phenomenological
EM form factor of the kaon (cf. Eq. (\ref{kff:vmd}))
is used the numbers in the second line of Table \ref{kkbar:tab2}
are reduced by 20\%.}  
\begin{table}
\begin{center}~
\begin{tabular}{|c||c|c|c|}\hline
Scenario & ${\rho_E^{(I=0)}}$ & ${\rho_M^{(I=0)}}$ & $\kappa^{(I=0)}$
\\\hline\hline
Exp't \cite{MMD96} & $-4.55$ & $-3.97$ & $-0.06$ \\\hline
AC/GS & $-0.50$ & $-0.09$ & $0.21$ \\\hline\hline
$\phi$ pole \cite{HMD96} & 2.21 & 2.87 & 0.15 \\\hline
\end{tabular}
\vspace{0.5cm}
\caption{\label{kkbar:tab2} Leading moments of the isoscalar
EM nucleon form factors. First line gives \lq\lq experimental'' 
values from the dispersion analysis of Ref. \cite{MMD96}. Second 
line shows $K\bar{K}$ contribution in AC/GS scenario. 
Third line gives contribution of $\phi$ pole alone in the pole fits of
Ref. \cite{HMD96}.}
\end{center}
\end{table}
Evidently, the $q^2$-dependence of the isoscalar EM form factors
is determined by states other than $|K\bar{K}\rangle$. Its contribution to
$\kappa^{(I=0)}$, however, is considerable. The VMD analyses of 
Refs. \cite{Jaf89,HMD96,For96} indicate a significant contribution to 
the leading moments from the $\phi$. The results of the VMD treatment 
and our present study
are compatible only if other intermediate states having $t_0\leq m_\phi^2$
contain considerable $\phi$ strength. A preliminary consideration of 
this possibility is given in Ref. \cite{HaM98}, where the role of the
$\phi$ resonance in the $3\pi$ channel is discussed (see also Ref. 
\cite{RMH98x}).

\section{Other Contributions and Conclusions}
\label{sec:conc}
The role of continuum and resonance effects in the isoscalar EM
and strangeness form factors appears considerably more complicated
than in the case of the isovector EM form factors whose
spectral functions are dominated by a combination of
$\rho$ resonance and uncorrelated $\pi\pi$ continuum.
In the present study, we have continued our previous efforts
\cite{MHD97,HaM98,RMH98} to determine the corresponding picture
for the isoscalar EM and strange vector current spectral functions.
We have focused on the $K\bar{K}$ contribution for two reasons:(i) the
availability of scattering data afford us with the least model-dependent
determination of this contribution to all orders in the strong coupling, 
and (ii) the OZI rule has prompted a number of strange form factor 
calculations assuming this state to give the dominant contributions. 
We find that

\medskip
\noindent (a) the $K\bar{K}$ contribution to the isoscalar EM and
strangeness electric spectral functions is significantly enhanced by the 
presence of a $\phi$-resonance in $\bpp$ and $\fka$.

\medskip
\noindent (b) there exists no evidence for such a resonance in the 
$\bpm$ partial wave.

\medskip
\noindent (c) the resonance affects only the normalization of
$G_M^a$ but not its $q^2$-dependence.

\medskip
\noindent (d) results (a) and (b) can be reconciled with a simple 
$\phi$-resonance model of $N\bar{N}\to K\bar{K}$ if the vector and tensor 
$\phi NN$ couplings have roughly equal magnitudes and opposite signs.
We obtain a value for $G_{\phi N\bar{N}}^V$ in agreement with the
VMD analyses of the isoscalar EM form factors \cite{MMD96}. 
Our value for $G_{\phi N\bar{N}}^T$, however, is larger in magnitude.

\medskip
\noindent (e) the $K\bar{K}$ contribution to the magnetic radius 
$\rho_M^s$ is significantly smaller than the value obtained at 
${\cal O}(p^3)$ in ChPT.

\medskip
\noindent (f) the $K\bar{K}$ contribution to the sub-leading 
$q^2$-dependence of the isoscalar EM moments is small. 
\medskip

\noindent
The result (f) implies that consideration of other intermediate
states is essential to a proper description of the isoscalar EM
and strangeness spectral functions. In this respect, the calculations
of Refs. \cite{GeI97,Bar98} are suggestive, indicating the possibility of
cancellations between different contributions as successive higher-mass
intermediate states are included. Moreover, as discussed in Refs. 
\cite{HaM98,RMH98x}, contributions from light, multi-meson intermediate 
states may be just as large as that of the $K\bar{K}$ state. 
Nevertheless, our study of the
$K\bar{K}$ state provides several insights into the the treatment of
these contributions. In particular, we are able to understand the
connection between continuum and resonance contributions to the
isoscalar EM and strangeness form factors and to evaluate the credibility
of other approaches used in computing them. Indeed, perturbative 
calculations which truncate at ${\cal O}(g^2)$ omit 
what appears to be the governing physics of the spectral functions, namely 
rescattering and resonance effects. Consequently,
the ${\cal O}(p^3)$ ChPT computation of the strange magnetic radius -- 
though counterterm independent -- contains only the non-analytic 
contributions at ${\cal O}(g^2)$ and exceeds our upper bound for the 
magnetic radius by an order of magnitude. The higher-order rescattering 
corrections needed to render the ChPT prediction consistent with our 
bound are presumably contained in terms of ${\cal O}(p^4)$ or higher.
Similarly, we suspect that the pattern of cancellations 
obtained in the ${\cal O}(g^2)$ NRQM calculation of Ref. \cite{GeI97} 
will be significantly modified when rescattering and resonance effects 
are included.  

A computation to all orders in $g$ of the remaining contributions would
appear to be a daunting task. A few observations may simplify the problem,
however. First, unitarity arguments suggest that the important
structure in the spectral function lies below the two-nucleon threshold.
Contributions from states such as $\Lambda\bar{\Lambda}, \Lambda
\bar{\Lambda}\pi,\ldots$ whose thresholds $t_\lambda > 4\mns$ are
limited by unitarity bounds on the strong partial waves\footnote{Note 
that this is not the case in one-loop model calculations where unitarity 
is strongly violated \cite{MHD97}.} and are unlikely
to be significantly enhanced by resonance effects in the intermediate
state form factors. 

Second, we expect that the only important pionic
contributions arise via resonances, such as the $\omega(780)$ or
$\phi(1020)$. In ChPT, for example, the matrix element$\langle 3\pi |\bar{s}
\gamma_\mu s| 0 \rangle$ receives no non-analytic contributions at 
${\cal O}(p^7)$. Consequently, the $3\pi$ contribution to the spectral
function is small in the absence of resonant short distance effects. 
As discussed in Ref. \cite{HaM98}, such effects (e.g., $3\pi
\leftrightarrow\omega$ and 
$3\pi\leftrightarrow\rho\pi\leftrightarrow\phi$) may enhance the
$3\pi$ contribution up to the scale of the $K\bar{K}$ contribution. 
In fact, we 
speculate that the $\phi$ strength obtained in the VMD analyses of the 
isoscalar EM form factors arises primarily in the $3\pi$ channel. There
exists little evidence for significant coupling of higher-mass 
multi-pion states to  $[I^G (J^{PC}) = 0^- (1^{--})]$ resonances. We
thus expect their contributions to be no larger than the non-resonant 
part of the $K\bar{K}$ term. 

Third, states involving pions and strange mesons may generate important
contributions via the $\omega(1420)$, $\omega(1600)$ and $\phi(1680)$ 
resonances. A preliminary exploration of this possibility is given in Ref.
\cite{Bar98}. In the VMD fits of Refs. \cite{Jaf89,HMD96,For96}, inclusion
of a vector meson pole in this mass region is needed to obtain an acceptable
$\chi^2$. Since the flavor content of the vector mesons in this region 
is not known, the higher mass contributions to the strangeness 
form factors have been inferred from a priori assumptions about their 
large-$t$ behavior. A reasonable range for the strange moments avoiding
these assumptions has recently been given in Ref. \cite{RMH98x}.

A calculation of unitarity bounds for states having $t_\lambda \geq 4\mns$ 
is tractable. Data for $N\bar{N}\to 3\pi$ and $\pi N\to \pi\pi N$ may 
permit a model-independent determination of the $3\pi$ contribution to 
all orders in $g$. Whether or not a realistic treatment of the other 
multi-meson states can be carried out remains to be seen.

\section*{Acknowledgement}
We thank T. Cohen, D. Drechsel, N. Isgur, R.L. Jaffe, and U.-G. 
Mei{\ss}ner for useful discussions. HWH acknowledges the hospitality 
of the INT in Seattle where part of this work was carried out.
MJR-M was supported under U.S. Department of Energy grant
DE-FG06-90ER40561 and a National Science Foundation Young
Investigator Award. HWH was supported by the Natural Sciences
and Engineering Research Council of Canada.

\end{document}